\documentclass[10pt,twocolumn]{article} 
\usepackage{simpleConference}
\usepackage{times}
\usepackage{graphicx}
\usepackage{amssymb}
\usepackage{url,hyperref}

\usepackage[nocompress]{cite}
\usepackage{amsmath,amssymb,amsfonts}
\usepackage{algorithmic}
\usepackage{textcomp}
\usepackage{fancyhdr}
\usepackage{enumitem}

\newcommand{\ignore}[1]{}

\usepackage{comment}
\usepackage{color}
\usepackage[pass]{geometry}
\usepackage[normalem]{ulem}
\usepackage{adjustbox}
\usepackage{multirow}
\usepackage{gensymb}
\usepackage{footnote}
\usepackage{threeparttable}
\usepackage[table,xcdraw]{xcolor}
\usepackage{xcolor}
\usepackage{soul}
\newcommand{\revisioncolor}[1]{{\color{black}{#1}}}

\usepackage{enumitem}
\usepackage{paralist}
\usepackage{caption} 
\usepackage[caption=false,font=footnotesize,labelfont=sf,textfont=sf]{subfig}

\begin{document}

\title{RISC-NN: Use RISC, NOT CISC as\\Neural Network Hardware Infrastructure}

\author{Taoran~Xiang$^{1,2}$,
	Lunkai~Zhang$^{1,3}$,
	Shuqian~An$^2$,
	Xiaochun~Ye$^2$,
	Mingzhe~Zhang$^2$,
	Yanhuan~Liu$^2$,\\
	Mingyu~Yan$^2$,
	Da~Wang$^2$,
	Hao~Zhang$^2$,
	Wenming~Li$^2$,
	Ninghui~Sun$^2$,
	and~Dongrui~Fan$^{2,4}$
}
\maketitle
\footnotetext[1]{Taoran Xiang and Lunkai Zhang contribute equally to the article.}
\footnotetext[2]
{
	Taoran Xiang, Shuqian An, Xiaochun Ye, Mingzhe Zhang, Yanhuan Liu, Mingyu Yan, Da Wang, Hao Zhang, Wenming Li, Ninghui Sun, and Dongrui Fan are with the State Key Laboratory of Computer Architecture, Institute of Computing Technology,
	Chinese Academy of Sciences, Beijing 100190, China\protect\\
	E-mail: taoran.xtr@gmail.com \{anshuqian, yexiaochun, zhangmingzhe, liuyanhuan, yanmingyu, wangda, zhanghao, liwenming, sunninghui, fandr\}@ict.ac.cn
}
\footnotetext[3]{Lunkai Zhang contributed to this project when he was a postdoctoral researcher at the University of Chicago. He is now working at Intel.\protect\\
	E-mail: lunkai.zhang.1984@gmail.com}
\footnotetext[4]{Dongrui Fan is with School of Computer and Control Engineering, University of Chinese Academy of Sciences, Beijing 100190, China.	}

	\begin{abstract}
		Neural Networks (\emph{NN}) have been proven to be powerful tools to analyze Big Data. However, traditional CPUs cannot achieve the desired performance and/or energy efficiency for NN applications. Therefore, numerous \emph{NN} accelerators have been used or designed to meet these goals. These accelerators all fall into three categories: \emph{GPGPUs}, \emph{ASIC NN Accelerators} and \emph{CISC NN Accelerators}. \emph{GPGPUs} achieve general purpose and high computing throughput, but cannot provide desired energy efficiency because their stream architecture cannot achieve efficient data reuse required by \emph{NN} applications. \emph{ASIC NN Accelerators} achieve best performance or energy efficiency through advanced data reuse optimization, however, they only support limited \emph{NN} use cases. \emph{CISC NN Accelerators} aim to achieve both general purpose and high energy efficiency by decomposing \emph{NN} applications into multiple relatively simple matrix or vector \emph{CISC} instructions. Though \emph{CISC NN Accelerators} can achieve considerable smaller memory footprint than \emph{GPGPU} thus improve energy efficiency; they still fail to provide same level of data reuse optimization achieved by \emph{ASIC NN Accelerators} because of the inherited poor pragrammability of their \emph{CISC} architecture.
		
		We argue that, for \emph{NN} Accelerators, \emph{RISC} is a better design choice than \emph{CISC}, as is the case with general purpose processors. We propose \emph{RISC-NN}, a novel many-core \emph{RISC}-based \emph{NN} accelerator that achieves high expressiveness and high parallelism and features strong programmability and low control-hardware costs. We show that, \emph{RISC-NN} can implement all the necessary instructions of state-of-the-art \emph{CISC NN Accelerators}; in the meantime, \emph{RISC-NN} manages to achieve advanced optimization such as multiple-level data reuse and support for \emph{Sparse NN} applications which previously only existed in \emph{ASIC NN Accelerators}. Experiment results show that, \emph{RISC-NN} achieves on average 11.88$\times$ performance efficiency compared with state-of-the-art Nvidia TITAN Xp \emph{GPGPU} for various \emph{NN} applications. \emph{RISC-NN} also achieves on average 1.29$\times$, 8.37$\times$ and 21.71$\times$ performance efficiency over \emph{CISC}-based \emph{TPU} in \emph{CNN}, \emph{MLP} and \emph{LSTM} applications, respectively. 
		Finally, \emph{RISC-NN} can achieve additional 26.05\% performance improvement and 33.13\% energy reduction after applying pruning for \emph{Sparse NN} applications.
	\end{abstract}

\section{Introduction}\label{sec:intro}
The world now has entered the \emph{Big Data Era} with millions of terabytes of data generated each day.
Such a huge amount of data, together with the computing power enabled by advanced processing 
technologies, has spurred a wave of powerful neural network (\emph{NN}) applications that play major roles 
in visual recognition \cite{NN_Visual_Recog}, voice recognition \cite{NN_Voice_Recog}, autonomous vehicle \cite{NN_Auto_Vehicle}
and smart advertisement campaign \cite{NN_Advertisement}. Unfortunately, traditional CPU/GPGPU platforms cannot achieve 
the desired performance and/or energy efficiency when running neural network applications \cite{Diannao}. Therefore,
it is a natural evolution to create specialized \emph{NN} accelerators. As a matter of fact, \emph{NN} applications
are ideal for implementing in accelerators because most of the \emph{NN} applications are composed of multiple 
relatively simple matrix or vector operations \cite{Cambricon}\cite{TPU} that can be efficiently implemented in hardware.

A straightforward solution in this respect is to design a  \emph{CISC} (Complex Instruction Set Computers) \emph{NN} 
accelerator with specialized ISAs for vector and/or matrix operations. This is exactly the design solution 
of prior arts used in two state-of-the-art \emph{NN} accelerators, TPU \cite{TPU} and Cambricon \cite{Cambricon}. Both accelerators use a simple centralized control 
logic to decode the \emph{CISC} instructions and then drive large arithmetic logic arrays. 
This design solution largely reduces the area and/or energy used by control logics---for example, the 
control logic of TPU takes about 2\% of the total chip area. As a result, these accelerators can
(1) achieve higher performance by implementing more arithmetic logics in the spare area and (2) achieve 
higher energy efficiency by decreasing the portion of energy used by control logics.


We argue, however, that, though well-motivated, these \emph{CISC NN} accelerators inherit some shortcomings
of \emph{CISC}, specifically a lack of instruction-level parallelism, and poor programmability.
The former shortcoming causes underutilization of execution units, which in turn lowers the performance
and energy efficiency of \emph{CISC NN} accelerators. The latter shortcoming results in performance penalties 
and energy inefficiency in \emph{CISC NN} accelerators when executing important \emph{NN} applications 
like \emph{CNN} and \emph{Sparse NN}. Since these shortcomings are intrinsic, they cannot be 
easily solved within the \emph{CISC} architecture.

In this paper, we aim to overcome the aforementioned problems using \emph{RISC-NN}, a novel many-core \emph{NN} 
accelerator that combines the merits of \emph{RISC} (Reduced Instruction Set Computers) and dataflow architecture.
We show that, compared with \emph{CISC NN} accelerators, \emph{RISC-NN} has the following merits:
\begin{compactitem}
\item \textbf{High Expressiveness}. As will be shown in Section \ref{sec:results201907:cisc}, \emph{RISC-NN} supports all the necessary CISC instructions 
of two state-of-the-art \emph{CISC NN} accelerators \cite{TPU,Cambricon}. Thus, it is at 
least as expressive as \emph{CISC NN} accelerators.
\item \textbf{High Parallelism.} As will be shown in Section \ref{sec:scheme}, \emph{RISC-NN} maximizes performance by exploiting parallelism at the task, \emph{ExeBlock} and instruction levels. 
\item \textbf{Strong Programmability.} As will be shown in Sections \ref{sec:results201907:cnn}\&\ref{sec:results201907:sparse}, for applications such as \emph{CNN} and \emph{Sparse NN}, \emph{RISC-NN} manages to implement state-of-the-art optimizations which previously were only available in \emph{ASIC NN Accelerators} \cite{Eyeriss,Cambricon-X}. This fact proves that \emph{RISC-NN} offers unprecedentedly strong programmability among all the \emph{NN} accelerators.
\item \textbf{Low Control Hardware Cost.} As will be shown in Section \ref{sec:results201907:hardware}, we can minimize the control hardware cost by carefully choosing features such as \emph{Very-RISC}
ISA, \emph{SIMD} and \emph{decoupling data movement and computation}.
\end{compactitem}
\textbf{In summary, we believe that using \emph{RISC} instead of \emph{CISC} 
as the hardware infrastructure for \emph{NN} applications is the right way to go.}

The rest of the paper is organized as follows. Section \ref{sec:motivation} discusses the shortcomings of state-of-the-art \emph{NN} accelerators and proposes design principles of \emph{RISC-NN}. 
Section \ref{sec:scheme} presents the design of \emph{RISC-NN}. Sections \ref{sec:method}\&\ref{sec:results} present the experiment 
methodology and results. Section \ref{sec:discuss} discusses some additional topics. Section \ref{sec:conclude} concludes the paper.
\vspace{-5pt}

\section{Motivation} \label{sec:motivation}
In this section, we first summarize state-of-the-art \emph{NN} accelerators, and then presents the conclusion which motivates our proposed \emph{RISC-NN} architecture.\vspace{-5pt}

\subsection{Shortcomings of State-of-the-Art NN Accelerators}

\label{sec:motivation:state-of-the-art}
Many architectures have been proposed in order to accelerate \emph{NN} applications. Yet, they all fall into three categories:
\emph{GPGPUs}, \emph{Special Purpose NN Accelerators} and \emph{CISC General Purpose NN Accelerators}.

\subsubsection{GPGPUs.}\label{sec:motivation:state-of-the-art:gpu}
GPGPUs \cite{GPGPU} are the most widely used \emph{NN} accelerators, due to their strong parallelism in processing of vector and matrix. Also, additional support has been provided to make GPGPUs more capable to accelerate \emph{NN} applications---take Nvidia GPGPUs for example: on software side, \emph{cuDNN}, a special deep \emph{NN} library has been introduced; on hardware side, mixed precision calculation unit has been introduced to reduce the calculation overhead of \emph{NN} applications. On the other hand, however, being a stream accelerator, GPGPU cannot provide fine-grained data sharing among its computing cores. As a result, data reuse can only happen in shared cache and global memory, and this introduces non-trivial excessive energy consumption.

Note that, very recently, Nvidia also has integrated \emph{Tensor Cores}, which are specialized in tensor operations, in their \emph{Volta} and \emph{Turing} architecture. The notion and architecture of \emph{Tensor Cores} is largely the same with \emph{TPU} \cite{TPU}, which will be discussed in detail in Section \ref{sec:motivation:state-of-the-art:cisc}.

\subsubsection{ASIC NN Accelerators.}\label{sec:motivation:state-of-the-art:asic}
\emph{ASIC NN Accelerators} \cite{NeuFlow,C-Brain2016Song,Minerva2016Reagen,isaac2016shafiee,shen2017maximizing,venkataramani2017scaledeep,FlexFlow2017Lu,Kwon2018MEF,Song2018GAN,Cai2018VHA,Outlier-Aware2018,Diffy,Laconic2019,TIE2019,UCNN2018,PermDNN2018,Cnvlutin2016albericio,EIE2016Han} are special-purpose accelerators tailored for one or few \emph{NN} applications, and they achieve high performance and/or energy efficiency by utilizing the intrinsic characterizations of the target \emph{NN} applications. For example: 
\begin{compactitem}
\item \emph{ASIC NN Accelerators} for \emph{CNN} applications \cite{Eyeriss,FlexFlow2017Lu,Kwon2018MEF} maximize the data reuse of \emph{CNN}. As a result, they decrease the off-chip/on-chip memory traffic and achieve high energy efficiency. 
\item \emph{ASIC NN Accelerators} for \emph{Sparse NN} applications \cite{Cambricon-X,PermDNN2018,Cnvlutin2016albericio,EIE2016Han} choose to take advantage of the ineffectual neurons in \emph{Sparse NN} by not transferring or computing them. This also largely reduces energy consumption and boosts performance.
\end{compactitem}

\subsubsection{CISC NN Accelerators.} \label{sec:motivation:state-of-the-art:cisc}
Designers also come up with more general \emph{NN Accelerators} which support various \emph{NN} applications and achieve better performance and/or higher energy efficiency than \emph{GPGPUs}. Interestingly, the field of these general \emph{NN Accelerators} is dominated by \emph{CISC} architectures. Two state-of-the-art commercial ones (TPU \cite{TPU} and Cambricon \cite{Cambricon}) are both \emph{CISC}. Such a design choice is motivated by the fact that most \emph{NN} applications can be factorized into multiple simple matrix and/or vector operations. As a result, it is a straightforward task to implement these operations in CISC instructions. Since the instructions of \emph{CISC NN} accelerators are relatively few and simple (compared to those in \emph{CISC general-purpose processors}), their control hardware is in fact quite lightweight: in TPU, the control logic only takes up just 2\% of the chip area. However, these \emph{CISC NN} accelerators still inherit the other two drawbacks of \emph{CISC architecture}: 
\begin{compactitem}
    \item \emph{Poor Instruction-Level Parallelism.} The designers of \emph{CISC NN} accelerators have striven to improve their instruction-level parallelism. For example, both \emph{TPU} and \emph{Cambricon} enable instruction pipelining, and \emph{Cambricon} even achieves out-of-order execution. However, to our best knowledge, all \emph{CISC NN} accelerators fall short in another fundamental aspect: the ability to share execution units among multiple instructions. That is, for \emph{TPU} and \emph{Cambricon}, a single instruction occupies a whole execution unit (e.g., the Matrix Multiply Unit of \emph{TPU} and the Vector/Matrix Func. Units of \emph{Cambricon}). This incurs considerable underuse of execution units when executed instructions cannot fully use the execution unit. For example, more than half of the MACs of \emph{TPU} are unused in \emph{CNN1} applications \cite{TPU}. Finally, such a drawback causes a nontrivial performance and energy efficiency overhead for \emph{CISC NN} accelerators.
    \item \emph{Poor Programmability.} It is hard, if not impossible, for \emph{CISC NN} accelerators to achieve the same advanced optimization achieved in \emph{ASIC NN Accelerators}. We take \emph{CNN} and \emph{Sparse NN} applications as examples:
    \begin{compactitem}
        \item \emph{CNN}. \emph{CISC NN} accelerators cannot fulfill all these data reuse techniques in \emph{ASIC NN accelerators} because the reused data are usually across multiple \emph{CISC} instructions, and it is hard to specify complex data reuse among different \emph{CISC} instructions.
        \item \emph{Sparse NN}. In order to efficiently decompose \emph{NN} applications into simple and regular CISC instructions, \emph{CISC NN Accelerators} require these applications to be regular. This is not the case for \emph{Sparse NN}. As a result, \emph{CISC NN Accelerators} could only treat \emph{Sparse NN} applications as regular dense ones and thus are incapable of benefiting from the computation and memory traffic benefits of \emph{Sparse NN}. 
    \end{compactitem}
\end{compactitem}

\subsubsection{RISC-Based NN Accelerators}\label{sec:discuss:risc-cores}
Several researches \cite{PULP-NN, GAP-8} study the possibility of using existing RISC cores to accelerate \emph{NN} applications. The main focus of these researches is for embedded systems which has a tight power budget. The RISC cores used (e.g., PULP) are usually light-weighted and tightly-coupled (e.g., shared instruction and/or data cache \cite{PULP-NN}). Such an approach works well in small scale system but also has scalability issues. On the other hand, many schemes of \emph{RISC-NN} (e.g., Distributed Control, Dataflow and Data Stationary) are proposed to enhance the scalability of the accelerator. We believe these core \emph{RISC-NN} ideas are the keys to scale up future \emph{RISC}-based accelerators.

\vspace{-10pt}
\subsection{Our Proposal: RISC-NN} \label{sec:motivation:risc}

We can see that the drawbacks of \emph{CISC NN Accelerators} mentioned in Section \ref{sec:motivation:state-of-the-art}
are fundamental and cannot be easily overcame within \emph{CISC} architecture. Here we propose \emph{RISC-NN},
a many-core \emph{RISC} accelerator that overcomes these drawbacks without sacrificing performance and energy efficiency. 
Before we describe \emph{RISC-NN} architecture in detail, we first introduce the techniques used in \emph{RISC-NN} and the reasons of using them:
\begin{compactitem}
\item \textbf{Distributed Control} (Section \ref{sec:scheme:pe-arch}). Unlike \emph{CISC NN} accelerators that use a centralized control scheme,
\emph{RISC-NN} uses a distributed control scheme in which each PE has its own control unit and can make most of the control decisions independently. With this capability, we can enable fine-grained control and data reuse, which are important to achieving desired programmability.
\item \textbf{Very-RISC ISA} (Section \ref{sec:scheme:isa}). We acknowledge that having a independent control unit for every PE may considerably 
increase the control overhead for the chip area and energy efficiency. To counteract such control overhead, we use a very simple 
RISC ISA in every PE, one with only 11 instructions and two different addressing modes. Yet, this ISA is flexible enough to represent 
and optimize state-of-the-art \emph{NN} applications.
\item \textbf{Decoupling Execution and Data Transmission} (Section \ref{sec:scheme:exeblock-manage}). In \emph{RISC-NN}, the data transmission (e.g., memory 
load and storage) is completely decoupled from execution (i.e., calculation). That is, all the data needed are preloaded into the PE's local 
memory prior to the execution; therefore, the execution will not be suspended by memory accesses. Such a design boosts the utilization
of execution components (i.e., MACs) and simplifies the control hardware.
\item \textbf{SIMD} (Section \ref{sec:scheme:simd}). Most \emph{NN} applications will be executed with identical instructions for multiple times. 
In order to take advantage of such an application feature, each PE uses a \emph{SIMD} (single instruction multiple data) architecture
to reduce the control energy overhead per operation and thus improve energy efficiency.
\item \textbf{Multi-Level Parallelism.} To achieve high utilization of execution units, 
we exploit task-level, \emph{exeBlock}-level and instruction-level parallelism (Sections \ref{sec:scheme:overview},  \ref{sec:scheme:exeblock-manage} and \ref{sec:scheme:pipeline}, respectively) in \emph{RISC-NN}.
\item \textbf{Replacing Large RFs with SRAMs} (Sections \ref{sec:scheme:rams}). It is well-known that a large RF (register file) takes a considerable
amount of power and energy \cite{Eyeriss}. Here we replace the large RF with a number of SRAMs and use hardware and
software mechanisms to guarantee that this SRAM-based approach can achieve performance similar to that of the RF-based approach.
\item \textbf{Dataflow} (Section \ref{sec:scheme:overview}). To reduce off-chip memory data transmission, we use an \emph{exeBlock}-level dataflow technique to enable inter-PE data reuse across \emph{exeBlocks}.
\item \textbf{Data Stationary} (Section \ref{sec:scheme:data-reuse}). To further reduce data movement, we allow multiple \emph{exeBlocks}
on the same PE to share the same piece of physical data. This technique is sometimes referred as \emph{data 
stationary} \cite{Eyeriss}.
\end{compactitem}
\vspace{-10pt}

\section{RISC-NN Architecture} \label{sec:scheme}
In this section, we introduce in detail the architecture of \emph{RISC-NN} with an 
emphasis on how we improve the accelerator's performance, efficiency and flexibility.

\vspace{-5pt}
\subsection{Overview}\label{sec:scheme:overview}

\begin{figure} 
    \begin{minipage}[b]{0.44\linewidth}
        \centering
        \includegraphics[width=\linewidth]{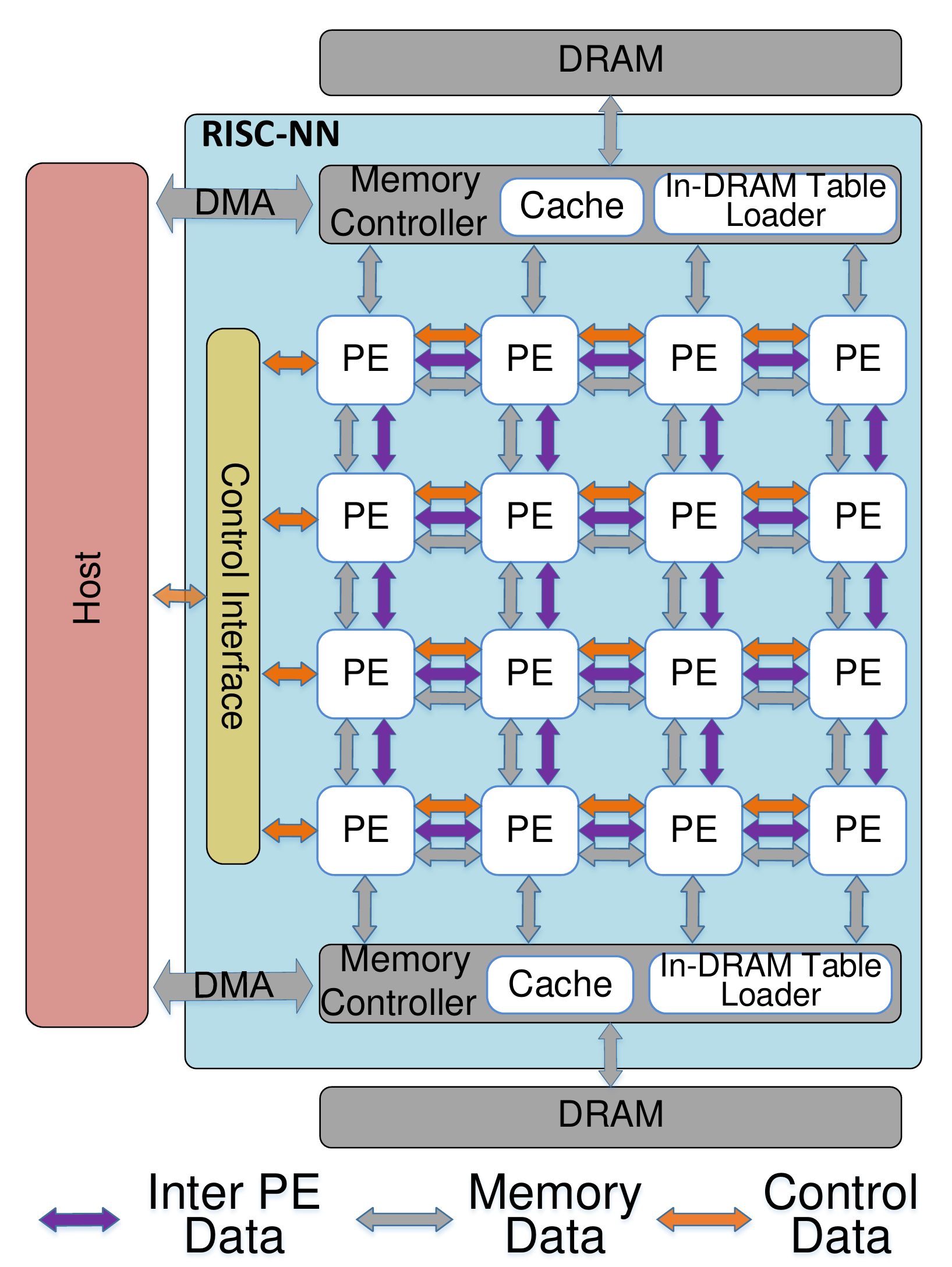}
	    \vspace{-0.22in}
        \caption{Overall Hardware Architecture of \emph{RISC-NN}.}
        \label{fig:scheme:hardware-arch}
    \end{minipage}
    \hspace{0.01\linewidth}
    \begin{minipage}[b]{0.48\linewidth}
        \centering
        \includegraphics[width=\linewidth]{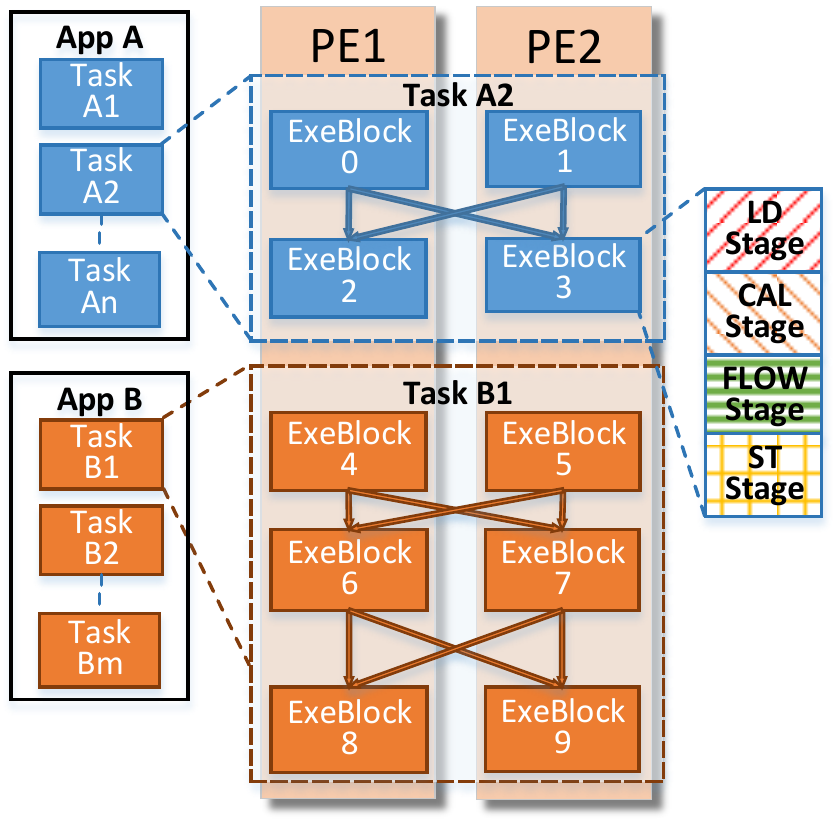}
        \vspace{15pt}
        \caption{Execution Model of \emph{RISC-NN}.}
        \label{fig:scheme:exec-model}
    \end{minipage}
    \vspace{-0.25in}
\end{figure}

Figure \ref{fig:scheme:hardware-arch} shows the overall architecture of
\emph{RISC-NN}. To reduce hardware complexity and achieve good scalability, 
\emph{RISC-NN} adopts a simple tiled architecture that consists of many homogeneous
PEs, several memory controllers with caches and \emph{In-DRAM Table Loader} modules (see Section \ref{sec:scheme:complex-funcs}) in their 
front end.
\emph{RISC-NN} talks to the host in two ways: control messages, which uploads the control information to PEs through a simple \emph{Control Interface}; and memory DMA, which transfers large amount of data (e.g., input data and instructions, output data) between host and DRAMs of \emph{RISC-NN}. Three sets of NoCs connect the components: (1) the Memory NoC, a MESH network
that transfers data and instructions between the off-chip DRAM and the PEs; (2) the Inter-PE NoC, a MESH
network that transfers data among different PEs; and (3) the Control NoC, a tree-like
network that connects the PEs with the \emph{Control Interface}. 

Before introducing the detailed PE architecture, we first introduce the execution
model of \emph{RISC-NN}. As is shown in Figure \ref{fig:scheme:exec-model},
multiple applications can run on \emph{RISC-NN} at the same time. \revisioncolor{Each application
consists of a sequence of consective tasks, and each task consists of multiple
execution blocks (i.e., \emph{ExeBlocks}) that are organized in a DataFlow manner.}
\emph{ExeBlocks} are then mapped into the PE arrays, with each PE containing multiple
\emph{ExeBlocks}. The code for each \emph{ExeBlocks} consists of up to four consective 
\emph{Execution Stages}, which we describe below:
\begin{compactitem}
\item \emph{LD (Load) Stage.} This stage loads data from the off-chip DRAM memory to the in-PE 
\emph{Operand RAM Module}. Once the associated task is enabled, an \emph{ExeBlock} can 
execute its \emph{LD Stage}.
\item \emph{CAL (Calculation) Stage.} This stage does all the calculation.  
An \emph{ExeBlock} can execute its \emph{CAL Stage} only when 
the following two conditions are met: first, its \emph{LD Stage} (if it exists) has 
already finished; second, it has received all the necessary data from its predecessor 
\emph{ExeBlocks} (i.e., finishing actication step, see Section \ref{sec:scheme:exeblock-manage}). 
\item \emph{FLOW (Flow) Stage.} This stage transfers data from the current \emph{ExeBlock}
to its successor.
\item \emph{ST (Store) Stage.} This stage transfers data from the in-PE \emph{Operand RAM Module} 
to the off-chip DRAM memory. As will be discussed in Section \ref{sec:scheme:complex-funcs}, in order to support complex activation/classifier functions, this stage can also do memory-side table lookup.
\end{compactitem}
\revisioncolor{In the beginning of an application, the host first uploads the instructions and data onto the DRAM through DMA. It then sends control messages to initialize the corresponding \emph{ExeBlocks} on the PEs. After that, as will be discussed in detail in section \ref{sec:scheme:exeblock-manage}, the PEs will automatically finish rest of the application execution such as instruction loading, task enabling, execution, etc., and store the final results in the DRAM for host to read back through DMA.}

\vspace{-5pt}
\subsection{Very-RISC ISA}\label{sec:scheme:isa}

\begin{table}
	\caption{\textbf{Very-RISC ISA of \emph{RISC-NN}.} All instructions have the same 
	             format: [OP F0, F1, F2, CTRL], in which \emph{OP} is 4-bit instruction type and \emph{F0}---\emph{F2} are 
	             three 16-bit operand fields. And \emph{CTRL} is a 12-bit control field which has a 8-bit \emph{Sparse PC Inc} sub-field to support \emph{Sparse NN} applications (see Section \ref{sec:results201907:sparse}) and a 4-bit \emph{In-DRAM Lookup Type} sub-field to support complex activation/classifier functions (see Section \ref{sec:scheme:complex-funcs}).
	             Each \emph{Execution Stage} has its own sub-instruction 
	             set. OPM is the abbreviation of \emph{Operand Memory} of the PE. DRAM 
	             refers to off-chip DRAM main memory. LD\_Base and ST\_Base refer to the \emph{Load
	             Base Address} and \emph{Store Base Address} of corresponding tasks, respectively.
	             }\label{table:scheme:isa}
	\centering
	\footnotesize
	\begin{threeparttable}
		\setlength\tabcolsep{1.5pt} 
		\begin{tabular}{|l|l|}
			\hline
			OP                 & Main Function                                           \\  \hline
			\multicolumn{2}{|c|}{\emph{LD Stage}}                                                                                      \\ \hline
			LD                 & OPM[F0]=DRAM[LD\_Base+\{F1,F2\}]       \\ \hline
			\multicolumn{2}{|c|}{\emph{CAL Stage}}                                                                        \\ \hline
			ADD               & OPM[F2]=OPM[F0]+OPM[F1]         \\ \cline{1-2}
			SUB               & OPM[F2]=OPM[F0]-OPM[F1]         \\ \cline{1-2}
			MUL               & OPM[F2]=OPM[F0]*OPM[F1]          \\ \cline{1-2}	
			MAX               & OPM[F2]=MAX(OPM[F0],OPM[F1])     \\ \cline{1-2}
			MIN               & OPM[F2]=MIN(OPM[F0],OPM[F1])     \\ \cline{1-2}
			MADD              & OPM[F2]=(OPM[F0]*OPM[F1])+OPM[F2]	\\ \cline{1-2}
			PREREAD0       & OP0-PreRead Data Reg = OPM[F0]     \\ \cline{1-2}
			PREREAD1       & OP1-PreRead Data Reg = OPM[F1]     \\ \hline
			\multicolumn{2}{|c|}{\emph{FLOW Stage}}                                                                      \\ \hline
			COPY             & PE[F2].OPM[F1]=OPM[F0]              \\ \hline
			\multicolumn{2}{|c|}{\emph{ST Stage}}                                                                                      \\ \hline
			ST                 & DRAM[ST\_Base+\{F1,F2\}]=OPM[F0]                                           \\ \hline	
		\end{tabular}
	\end{threeparttable}
	\vspace{-0.2in}
\end{table}

As discussed in Section \ref{sec:motivation:risc}, the ISA of \emph{RISC-NN} should be
expressive enough to represent and optimize state-of-the-art \emph{NN} applications 
but also simple enough to incur minimal area and power/energy overhead.
We have designed a Very-RISC ISA to achieve such a goal.

Table \ref{table:scheme:isa} summarizes \emph{RISC-NN} ISA, which includes
just 11 fixed-length instructions. All instructions have the same format:  [OP F0, F1, F2, CTRL], 
in which OP is 4-bit instruction type field, F0---F2 are three 16-bit
operand fields and CTRL is a 12-bit control field with a 8-bit \emph{Sparse PC Inc} sub-field to support \emph{Sparse NN} applications (see Section \ref{sec:results201907:sparse}) and a 4-bit \emph{In-DRAM Lookup Type} sub-field to support complex activation/classifier functions (see Section \ref{sec:scheme:complex-funcs}).  Each instruction 
only belongs to one \emph{Execution Stage}. The ISA contains only two addressing modes:
\begin{compactitem}
\item \emph{Direct PE Addressing Mode.} For all the in-PE data, we
use a simple direct addressing mode: one 16-bit field representing the data's absolute address inside the
\emph{Operand RAM Module} of the PE. If the instruction has a remote PE address (i.e., \emph{COPY} 
instruction), then another 16-bit field represents the remote PE number.
\item \emph{Base-Plus-Offset DRAM Addressing Mode.} To access data in the off-chip DRAM,
we use a base-plus-offset addressing mode in which the DRAM address of a piece of data is the sum of the following
two components: (1) a pre-stored 32-bit base address of the corresponding task and (2) a 32-bit offset address
represented by two 16-bit fields of the instruction (i.e., F1\&F2). 
\end{compactitem}
Here we describe in detail the individual instructions:
\begin{compactitem}
\item \emph{LD and ST Instructions.} These two instructions are in charge of
transferring data between the DRAM and the PE's local \emph{Operand Memory}.
They adopt the \emph{Direct PE Addressing } and \emph{Base-Plus-Offset 
DRAM Addressing Modes}. Note that \emph{LD} and \emph{SD} instructions
use different pre-stored base addresses (i.e., \emph{LD\_Base} and \emph{ST\_Base}).
As discussed in Section \ref{sec:scheme:data-reuse}, we can use \emph{ST}
and \emph{LD} instructions to conveniently share data between different tasks or applications. Section \ref{sec:scheme:complex-funcs} also shows that, with non-zero \emph{In-DRAM Lookup Type} bits in \emph{CTRL} field, \emph{ST} supports complex activation/classifier functions. 

\item \emph{CAL Instructions.} There are eight \emph{CAL} instructions. All of them 
use the \emph{Direct PE Addressing Mode} since they can only visit the data in the PE's 
local \emph{Operand Memory}. Six out of eight instructions (i.e., \emph{ADD, SUB, MUL, MAX, MIN, 
MADD}) are calculation-style instructions. The rest, the \emph{PREREADs}, are used to avoid Operand RAM
Module access conflicts (see Section \ref{sec:scheme:control-bits}). 

\item \emph{FLOW Instruction.} The \emph{COPY} instruction transfers data between the 
\emph{Operand RAM Modules} of source and destinatoin PEs. It enables data sharing among
\emph{ExeBlocks} of the PEs in a dataflow style.
\end{compactitem}

\vspace{-5pt}
\subsection{Overall Architecture of PE}\label{sec:scheme:pe-arch}

\begin{figure}
\centering
\vspace{-0.08in}
\includegraphics[width=\linewidth]{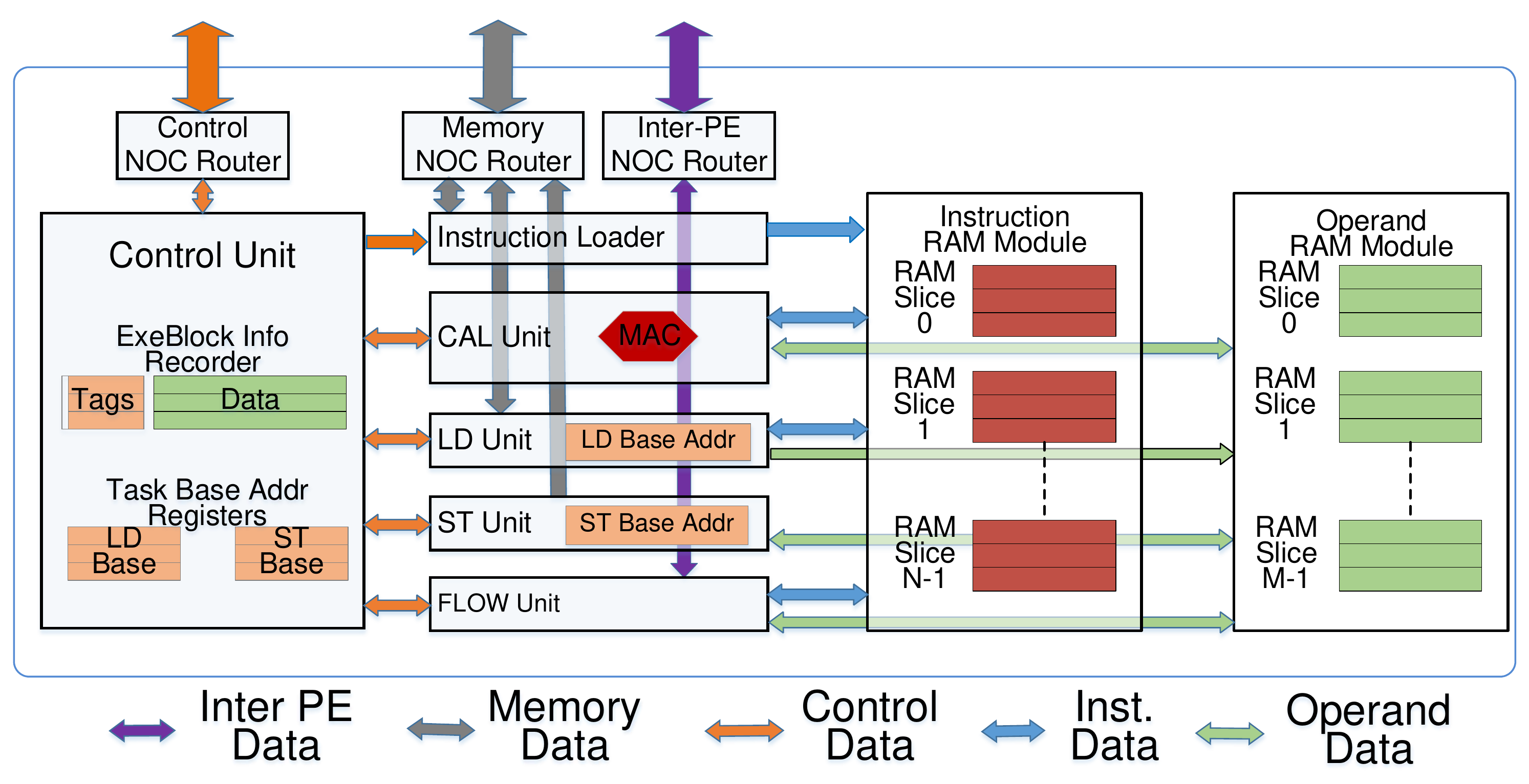}
\vspace{-0.23in}
\caption{PE Architecture of \emph{RISC-NN}.} \label{fig:scheme:pe-arch}	
\vspace{-0.25in}
\end{figure}

We carefully designed the PE architecture of \emph{RISC-NN} based on its execution model.
As shown in Figure \ref{fig:scheme:pe-arch}, \emph{RISC-NN} PE consists of the following
modules: 
\begin{compactitem}
\item \emph{Control Unit}. This is the central control logic of the PE. It receives and records the 
control information of the corresponding \emph{ExeBlocks} and \emph{tasks} and manages the 
operations of \emph{Execution Units} and the \emph{Instruction Loader}. 

\item \emph{Execution Units} (\emph{LD/CAL/FLOW/ST Units}). Each \emph{ExeBlock} stage
has its own \emph{Execution Unit}. As shown in Section \ref{sec:scheme:exeblock-manage}, 
such a decoupling method is the key to improving the component's utilization.

\item \emph{Instruction Loader.} This unit loads the instructions of \emph{ExeBlocks} from the off-chip
memory to the in-PE \emph{Instruction Memory}. It updates \emph{Sparse PC Inc} bits in control field when running \emph{Sparse NN} applications.

\item \emph{Operand and Instruction RAM Modules.} These in-PE RAM modules store operands 
(i.e., data) and instructions of the corresponding \emph{ExeBlocks}, respectively. Both modules consist
of multiple uniformly addressed RAMs. Section \ref{sec:scheme:rams} will discuss their architecture details.

\item \emph{Routers.} Three routers connect the PE to the Memory, Inter-PE and Control NoCs,
respectively.
\end{compactitem}

\vspace{-10pt}
\subsection{Management of ExeBlocks}\label{sec:scheme:exeblock-manage}

\ignore{
\begin{figure*}[h]
    \begin{minipage}[b]{0.485\linewidth}
        \centering
        \includegraphics[width=\linewidth]{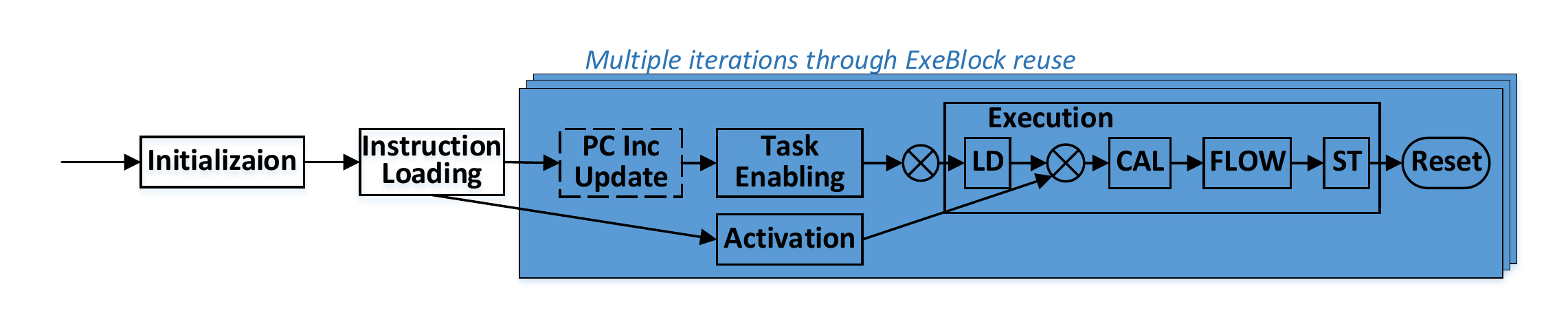}
	  \vspace{-0.25in}
        \caption{The Execution Order of Steps and Stages of an \emph{ExeBlock}.}
        \label{fig:scheme:step-stage-order}
    \end{minipage}
    \hspace{0.02\linewidth}
    \begin{minipage}[b]{0.435\linewidth}
        \centering
        \includegraphics[width=\linewidth]{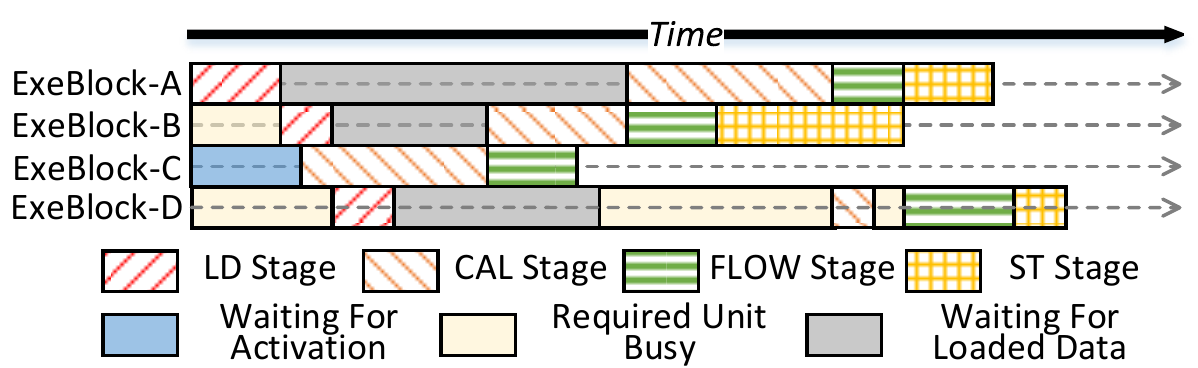}
	  \vspace{-0.2in}
        \caption{Overlapping the Steps and Stages of an \emph{ExeBlock}.}
        \label{fig:scheme:step-stage-overlapping}
    \end{minipage}
\end{figure*}
}

The instructions on a PE are managed by the \emph{Control Unit} at the granularity of \emph{ExeBlock}.
As shown in Figure \ref{fig:scheme:pe-arch}, an \emph{ExeBlock Info Recorder} in the 
\emph{Control Unit} records all the necessary information of its \emph{ExeBlocks}. 
An \emph{ExeBlock} is then managed in a multi-step/stage manner. 
Figure \ref{fig:scheme:step-stage-order} summarizes the order of steps and stages of 
an \emph{ExeBlock}. Here we introduce in detail the steps and stages:
\begin{compactitem}

\item \textbf{Initialization Step.} This step initializes a new \emph{ExeBlock}. It specifies the
following information of the \emph{ExeBlock}: 
\begin{compactitem}
\item \emph{Priority}, which records its scheduling priority. 
\item \emph{Task ID}, which records the corresponding Task ID of the corresponding task.
\item \emph{\#Predecessor}, which records the number of predecessor \emph{ExeBlocks}.
\item \emph{Stage Starting PCs}, each of which records the starting PC of the corresponding \emph{ExeBlock Stage}.
\item \emph{Stage Ending PCs}, each of which records the ending PC of corresponding \emph{ExeBlock Stage}. If the Starting and
Ending PCs of an \emph{ExeBlock Stage} are the same, it means the \emph{ExeBlock} does not have this stage.
\item \emph{Inst DRAM Address}, which records the starting DRAM address of \emph{ExeBlock} instructions.
\item \emph{Successor ExeBlocks}, which points to all the succesor \emph{ExeBlocks} of current \emph{ExeBlock}. In our current implementation, each \emph{ExeBlock} has up to 3 successors.
\item \emph{Sparse Execution}, this bit indicates if this \emph{ExeBlock} is used for a \emph{Sparse NN} application or not.
\end{compactitem}

\item \textbf{Instruction Loading Step.} When \emph{Instruction Loader} is not busy,
it finds the \emph{ExeBlock} with the highest priority among the ones whose instructions have not been loaded yet and then loads the instructions from the corresponding \emph{Inst DRAM 
Address} into the corresponding location of the \emph{Instruction RAM Module}. Note that, we store all the instructions 
of an \emph{ExeBlock} consecutively in the \emph{Instruction RAM Module}. As a result we can load them 
at once in a DMA manner.

\item \textbf{Sparse PC Inc Update Step.} Only \emph{ExeBlocks} with \emph{Sparse Execution} have this step. At the beginning of each \emph{Sparse NN} task, the \emph{Control Interface} sends a sparse vector to each \emph{ExeBlocks}. For each \emph{ExeBlock}, the bits of its sparse vector is the same with its instruction count, with each bit indicating if the instruction should be executed in the upcoming task. \emph{Instruction Loader} takes the sparse vector and translates it into \emph{Sparse PC Inc} of each individual instruction, which is the PC increment to the next valid instruction, and write such information to the \emph{Inst RAM Module}.

\item \textbf{Task-Enabling Step.} 
\emph{Control Interface} enables the task by broadcasting a 
\emph{task-enabling message} to all the PEs. An \emph{ExeBlock} must have its corresponding 
task enabled to start its \emph{Execution Step.}

\item \textbf{Activation Step.} At the end of the \emph{FLOW Stage} of an \emph{Execution Step}, 
an \emph{ExeBlock} sends \emph{activation} requests to all its successor \emph{ExeBlocks}, 
indicating that all the necessary data have been transferred. The \emph{Activation Step} 
finishes only when the \emph{ExeBlock} successfully collects \emph{activation} requests 
from all its predecessors.

\item \textbf{Execution Step.} This is the main step, which executes the instructions of the \emph{ExeBlock}.	
As mentioned in Section \ref{sec:scheme:overview}, this step contains four consecutive stages: 
\emph{LD}, \emph{CAL}, \emph{FLOW} and \emph{ST}. Note the following: 
\begin{compactitem}
\item An \emph{ExeBlock} can start \emph{Execution Step} only when \emph{Instruction 
Loading} and \emph{Task-Enabling Steps} have finished. 
\item A stage of \emph{Execution Step} can start only when all its previous stages were 
executed or do not exist.
\item In order to collect all the necessary data, the \emph{CAL} stage of \emph{Execution Step} 
can start only when the \emph{Activation Step} has finished.

\end{compactitem}

\item \textbf{Reset Step.} After the \emph{Execution Step}, the \emph{Control Unit} resets the status of 
the \emph{ExeBlock} before its termination. Such an operation is necessary for
the reuse of an \emph{ExeBlock} (see Section \ref{sec:scheme:data-reuse}).

\end{compactitem}

As shown in Figure \ref{fig:scheme:step-stage-overlapping}, such a multi-step/stage management method overlaps the steps and stages of 
different \emph{ExeBlocks}. This in turn helps maximize the utilization of the PE execution units, especially \emph{CAL Units}. This is important because improving the utilization of \emph{CAL Units} is the key to improving the 
performance and energy efficiency of the whole \emph{RISC-NN}.

\vspace{-5pt}
\subsection{Instruction and Operand RAM Modules}\label{sec:scheme:rams}

\begin{figure}
    \centering
    \vspace{-0.1in}
    \includegraphics[width=\linewidth]{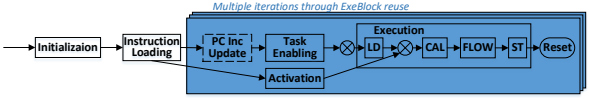}
    \vspace{-0.3in}
    \caption{The Execution Order of Steps and Stages of an \emph{ExeBlock}.}
    \label{fig:scheme:step-stage-order}
    \vspace{-0.13in}
\end{figure}

\begin{figure}
    \centering
    \vspace{-0.13in}
    \includegraphics[width=0.8\linewidth]{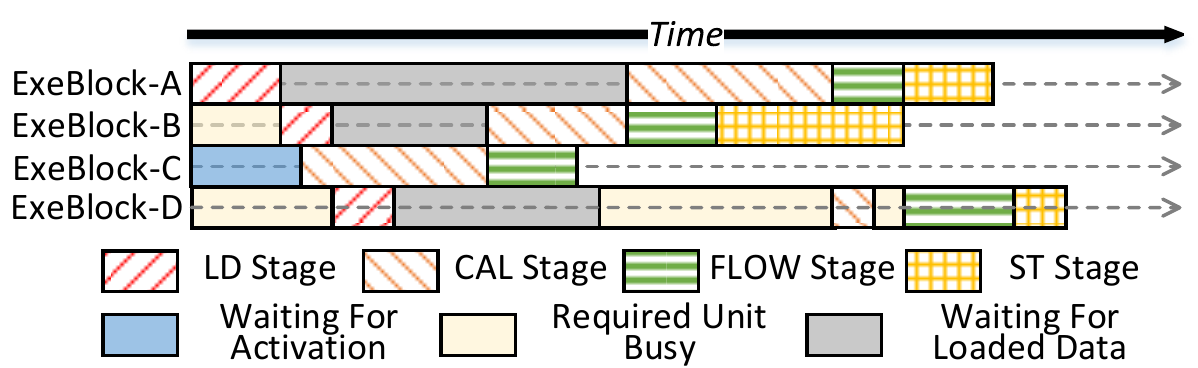}
    \vspace{-0.05in}
    \caption{Overlapping the Steps and Stages of an \emph{ExeBlock}.}
    \label{fig:scheme:step-stage-overlapping}
    \vspace{-0.2in}
\end{figure}

This section discusses the hardware structures used to store instructions and operands (i.e., data) in \emph{RISC-NN}. These are the \emph{Operand} and \emph{Instruction RAM Modules}, respectively.

\begin{figure} 
\centering
\vspace{-0.03in}
\includegraphics[width=0.97\linewidth]{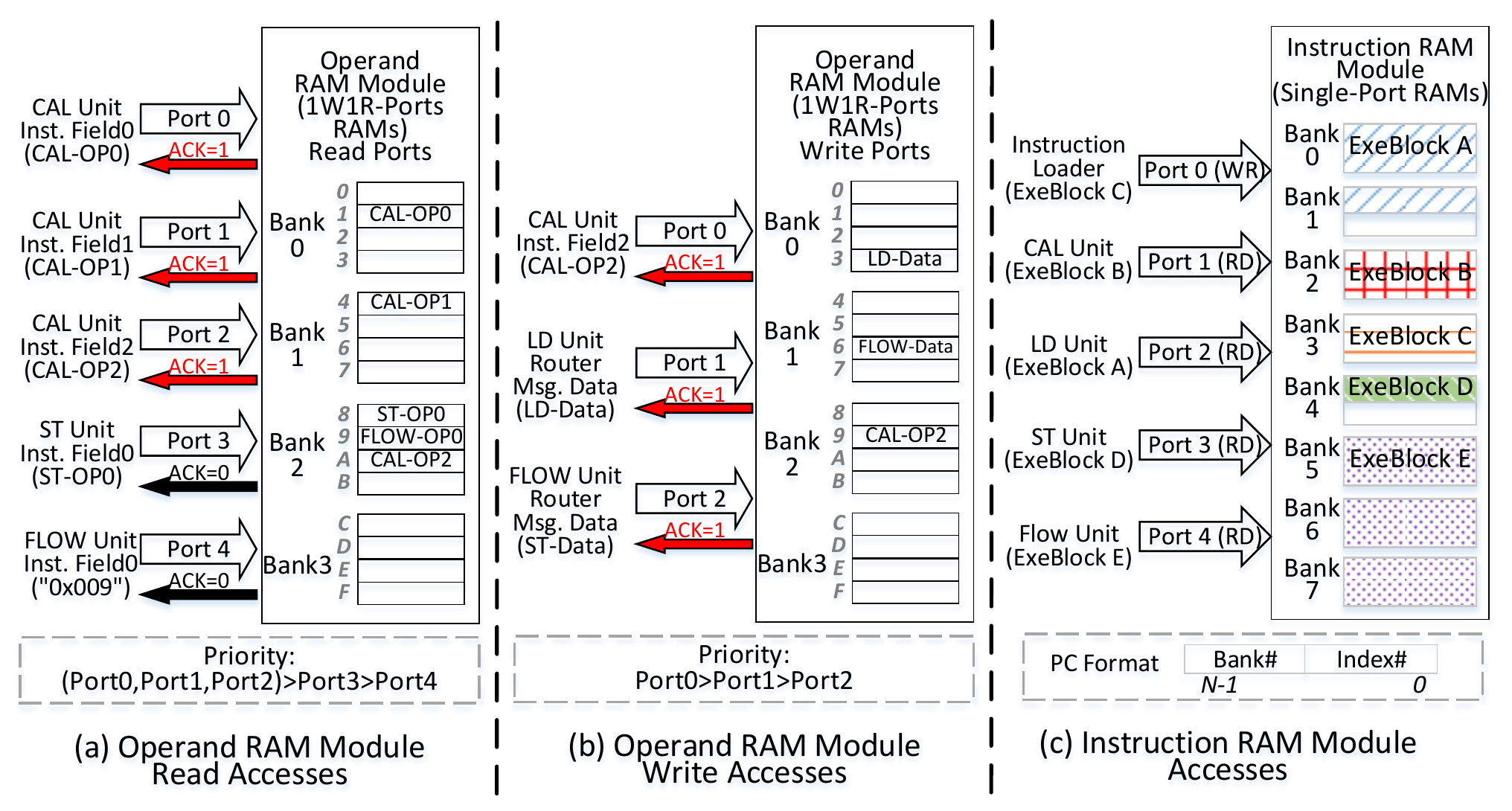}
	\vspace{-0.1in}
\caption{Concept of Instruction and Operand RAM Modules.}
\label{fig:scheme:pe-memory}

\end{figure}

\textbf{\emph{Instruction RAM Module.}} As shown in Figure \ref{fig:scheme:pe-memory} (c), the \emph{Instruction RAM Module} consists of multiple (8 in our experiment) single-port SRAM banks. Each bank can only be occupied by a single \emph{Execution Unit} at any time.


\textbf{\emph{Operand RAM Module.}} One important reason for why general-purpose architectures (e.g., GPU and CPU) are not energy efficient enough is that they extensively use large, power-hungry RFs. To improve energy efficiency, we replace the RF with a low-cost \emph{Operand RAM Module}, which can be conceptually viewed as a large multi-write-multi-read SRAM. The \emph{Operand RAM Module} consists of multiple (16 in our experiment) 1-write-1-read SRAM banks, each of which can serve at most one write and one read at the same time. As shown in Figure \ref{fig:scheme:pe-memory} (a), three out of five read ports (Ports 0--2) are used by the \emph{CAL Unit}. These three ports have the highest priority and must be served at the same time. The ports of \emph{ST} and \emph{Flow Units} have a lower priority. As a result, \emph{CAL Instructions} can always be served without stalling (i.e., ack always be 1'b1), but \emph{ST} and \emph{Flow Instructions} may have to wait because of the bank conflicts. To avoid the conflict of \emph{CAL Unit} read ports, we map input operands of a \emph{CAL} instruction to be distributed in different banks during compilation. As we discuss in Section \ref{sec:scheme:control-bits}, if there are conflicts remain in a \emph{CAL Instruction}, we use two pre-read instructions (i.e., PREREAD0/1) to resolve them.

\vspace{-10pt}
\subsection{Four-Stage CAL Unit Pipeline}\label{sec:scheme:pipeline}

\begin{figure} 
\centering
\vspace{-0.2in}
\includegraphics[width=\linewidth]{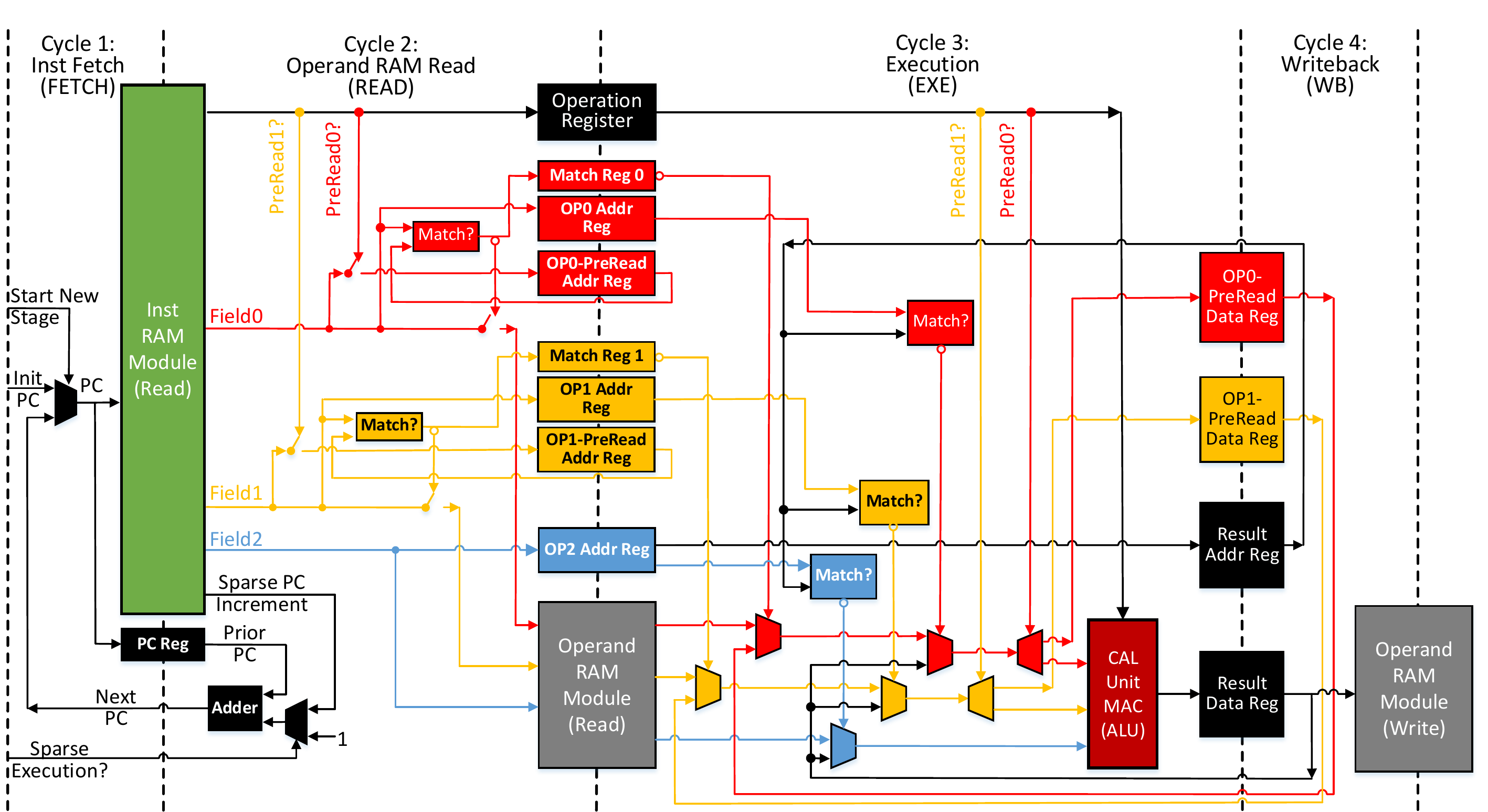}
\vspace{-0.25in}
\caption{Execution Pipeline of \emph{CAL Unit}.}
\label{fig:scheme:pipeline}
\vspace{-0.15in}
\end{figure} 

When compared to a conventional five stage \emph{RISC} pipeline, the pipeline of the \emph{CAL Unit} has 
several differences:
\begin{compactitem}
\item There is no need to have one separate \emph{Instruction Decode} stage because the ISA is 
ultra simple and thus decoding takes a very short time. 
\item the \emph{Memory Access} stage can also be omitted since there is no memory access in the \emph{CAL Unit}. 
\item One extra \emph{Operand RAM Read} (READ) stage is needed to get access to the data in \emph{Operand RAM Module}. 
\end{compactitem}
Therefore, as shown in Figure \ref{fig:scheme:pipeline}, the \emph{CAL Unit} uses a short four-stage (FETCH, READ, EXE and Writeback) pipeline.

As is mentioned in Section \ref{sec:scheme:exeblock-manage}, all the instructions are stored consecutively in \emph{Instruction RAM Module}. Therefore, for regular (\emph{non-sparse}) \emph{NN} applications, the pipeline just increases its current PC by 1 to fetch the next instruction. For \emph{sparse NN} applications, the pipeline uses the value of \emph{Sparse PC Inc} (see Sections \ref{sec:scheme:isa} \& \ref{sec:scheme:exeblock-manage}) in control field of current instruction to get the next instruction.

\vspace{-10pt}
\subsection{Resolving Data Hazard and Operand Conflicts}\label{sec:scheme:control-bits}


\begin{figure} [t]
\centering
\includegraphics[width=0.90\linewidth]{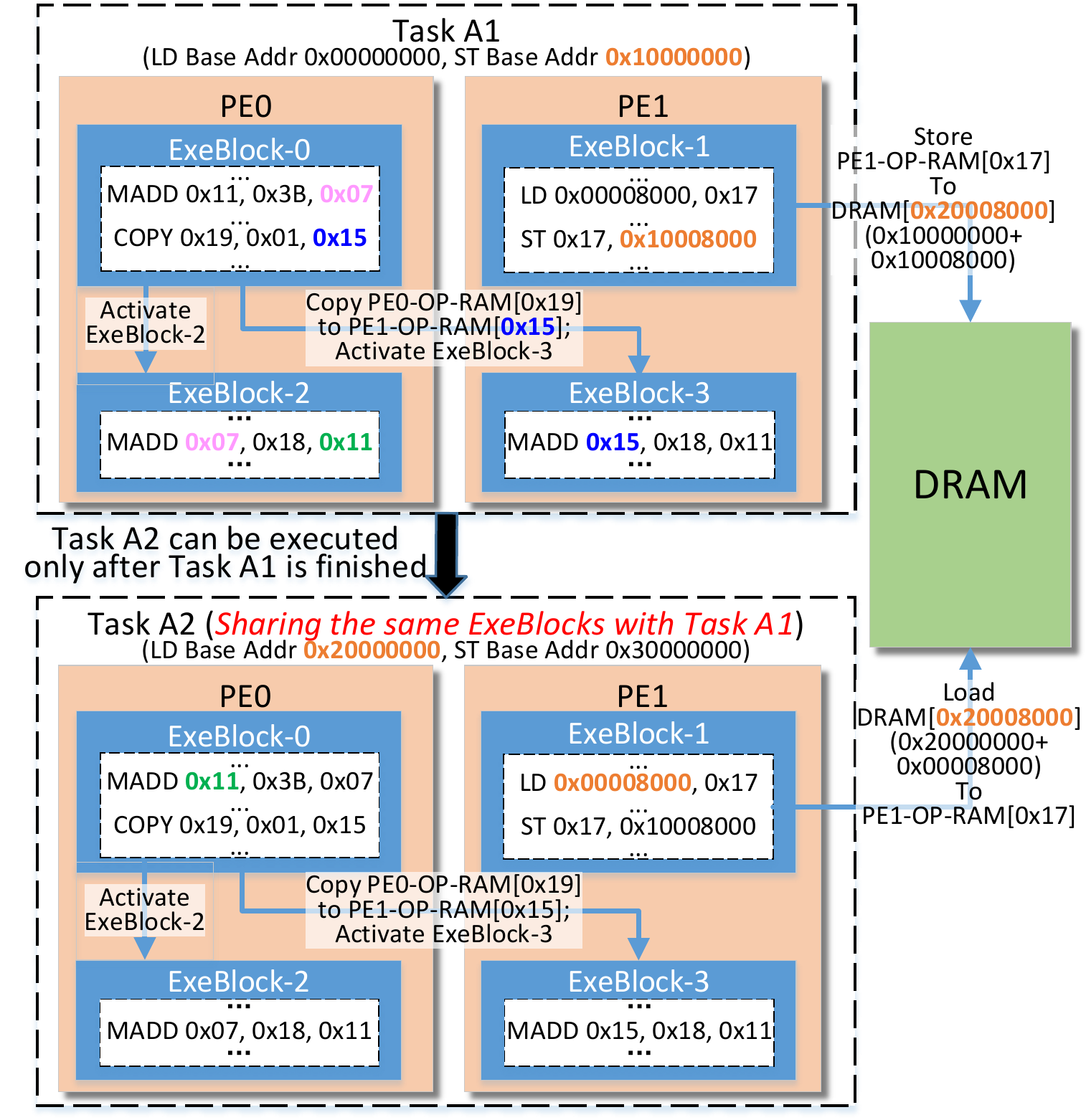}
\vspace{-0.05in}
\caption{ExeBlock Reuse and Sharing.}\label{fig:scheme:data-reuse-sharing}
	\vspace{-0.3in}
\end{figure} 

\begin{table*}[t]
	\centering
	\footnotesize
	\caption{\emph{RISC-NN Simulation Parameters}}\label{table:method:parameters}
\begin{tabular}{|l|l|}
\hline
\#PEs      & 64                                                                                         \\ \hline
Memory    & \revisioncolor{1 DDR4 2400MHz Controller with 1MB write-back policy front-end cache, which is distributed into 8 slices.}                                        \\
Subsystem & \revisioncolor{Each cache slice has 4 ways and 64-byte block size, and is connected to an edge memory NoC router.} \\ \hline
PE        & SIMD-8, 16-bit MAC Unit; 8-bank Instruction RAM Module, with each bank using a single 64bit*512 single port RAM;   \\
          & 16-bank Operand RAM Module, with each bank using a single 128bit*128 1W1R RAM;             \\
          & 85-bit width Control NoC Router; 128-bit data width Memory \& Inter-PE NoC Routers.        \\ \hline
\revisioncolor{Host Access} & \revisioncolor{PCIE3.1\cite{Pciepower}; 5mW/Gb/lane} \\ \hline
\revisioncolor{Frequency}      & \revisioncolor{1.887GHz, which results in a peak performance of 1.932TOPS ($1.887G\times64\times8\times2$, a MAC unit delivers 2OPs per cycle)}       \\ \hline
\end{tabular}
	  \vspace{-0.2in}
\end{table*}

Though we omit memory accesses and jump instructions, there are still two more situations in the \emph{CAL Unit} needing special attention: (1) \emph{Operand RAM Conflict}, in which two or more operands of an instruction come from the same \emph{Operand RAM} bank so they cannot be read at the same time and (2) \emph{Read-After-Write Data Hazard}, in which an instruction uses the result of the instruction right before it.

\textbf{Resolving Operand RAM Conflict}. As is shown in Table \ref{table:scheme:isa} and Figure \ref{fig:scheme:pipeline}, we have two pre-read instructions (i.e., PREREAD0/1) which read the data from \emph{Operand RAM} into OP0/1-PreRead Data Reg and store the corresponding \emph{Operand RAM} addresses in \emph{OP0/1-PreRead Addr Reg}. If the compiler detects \emph{Operand RAM Conflict} in one instruction, it injects pre-read instruction(s) up front. At READ stage, if \emph{CAL Unit} detects that the address of an OP (i.e., \emph{OP0}) is the same with its pre-read address register (i.e., \emph{OP0-PreRead Addr Reg}), it will bypass reading \emph{Operand RAM} and use pre-read data (i.e., from \emph{OP0-PreRead Data Reg}) instead. Note that, data in pre-read address \& data registers are one-time only---once used, they will be invalidated.

\textbf{Resolving Read-After-Write Data Hazard}. As is shown in Figure \ref{fig:scheme:pipeline}, at EXE stage, \emph{CAL Unit} compares the address of each operand (i.e., \emph{OP0/1/2 Addr Reg}) with the address of result in previous cycle (i.e., \emph{Result Addr Reg}). If any of them match, \emph{CAL Unit} then uses the result of previous cycle, which is stored in \emph{Result Data Reg}.

\vspace{-0.15in}
\subsection{SIMD}\label{sec:scheme:simd}

Most \emph{NN} applications are static and need to be executed multiple times (this feature is sometimes referred as a \emph{Single-Graph 
Multiple Flow} \cite{Single-Graph-Multiple-Flows}). Therefore they are a perfect match for \emph{SIMD} \cite{SIMD} execution. As a result, our \emph{RISC-NN}
also adopts \emph{SIMD} (\emph{SIMD-8} in our experiment) to improve performance and reduce control overhead.

\vspace{-0.1in}
\subsection{Supporting Complex Activation \& Classifier Functions}\label{sec:scheme:complex-funcs}

For design complexity and hard-ware cost considerations, we choose to keep the ISA of \emph{RISC-NN} simple, which does not support complex activation/classifier functions. Instead, we rely on table lookup \cite{Diannao} to implement these complex functions. We also have the following two observations:
\begin{compactitem}
\item For 16-bit accuracy (i.e., accuracy of \emph{RISC-NN}), we need a table of $2^{16}$ entries to achieve full accuracy lookup. This corresponds to 128KB of memory space, which might be too large for on-chip SRAM but is trivial for off-chip memory. 
\item These functions are usually used at the very end of the layer computation and comprise a very small fraction of the overall computation.
\end{compactitem}
Based on the above observations, we choose to store the lookup tables of complex functions in the DRAM with each table taking only 128KB of DRAM space. As is discussed in Section \ref{sec:scheme:isa}, each \emph{ST} instruction has 4-bit \emph{In-DRAM Lookup Type} in its CTRL field. Non-zero value of this type means there is a complex function lookup associated with the stored value. In that case, the memory controller first uses \emph{In-DRAM Table Loader} (See Figure \ref{fig:scheme:hardware-arch}) to look up the function value in corresponding in-DRAM table and then store this value into DRAM/cache instead.

\vspace{-0.1in}
\revisioncolor{
\subsection{Cache Architecture}
Although \emph{RISC-NN} strives to reduce memory access by utilizing intra-PE and inter-PE data reuse, sometimes there are still considerable amount of memory accesses to large shared global data structures (such as feature map in \emph{CNN}). To enhance the performance and energy efficiency of these memory accesses, we introduce a cache module as the front-end of memory controller. This is a write-back cache which accommodates all the LD and ST requests, but bypasses all the instruction loading requests.
}

\vspace{-0.1in}
\subsection{Data Reuse and Sharing}\label{sec:scheme:data-reuse}
As shown in Figure \ref{fig:scheme:data-reuse-sharing}, \emph{RISC-NN} supports various data reuse and sharing mechanisms that are 
important to improving its performance and energy efficiency. We describe these mechanisms below:
\begin{compactitem}
\item \emph{ExeBlock Reuse.} Tasks A1 and A2 share the same \emph{ExeBlocks}. As a result, the instructions
of the \emph{ExeBlocks} only need to be loaded once at the beginning of Task A1.
\item \emph{Inter-ExeBlock Data Reuse.} \emph{ExeBlock-0} and \emph{ExeBlock-2} of Task A1 can get access
to the same operand in index \textbf{0x07} of PE0's \emph{Operand RAM Module}.
\item \emph{Inter-Task Data Reuse.} Both \emph{ExeBlock-2} of Task A1 and \emph{ExeBlock-0} of Task A2 
can get access to the operand in index \textbf{0x11} of PE0's \emph{Operand RAM Module}.
\item \emph{Inter-PE Data Sharing.} During \emph{FLOW Stage}, \emph{ExeBlock-0} copies the operand
in index \textbf{0x11} of PE0's \emph{Operand RAM Module} to index \textbf{0x15}
of PE1's \emph{Operand RAM Module}. Since \emph{ExeBlock-0} is a 
predecessor of \emph{ExeBlock-3},  it has to send a network message to activate \emph{ExeBlock-3} at the end 
of its \emph{FLOW Stage}.
\item \emph{In-Memory/Cache Data Sharing.} Though it is best to share data through \emph{Inter-ExeBlock Data Reuse}
and \emph{Inter-Task Data Reuse}, sometimes the amount of data that needs to be shared exceeds the capacity of
the \emph{Operand RAM Module}. In this case, we need to share data through the main memory/cache. As shown in the
figure, by assigning a different \emph{LD Base Addr} and \emph{ST Base Addr}, Tasks A1 and A2 manage to share
data by storing to and loading from the same memory address \textbf{0x20008000}.
\end{compactitem}


\vspace{-0.1in}
\subsection{Programming Model}\label{sec:scheme:programming-model}

\begin{figure} 
\centering
\vspace{-0.1in}
\includegraphics[width=0.9\linewidth]{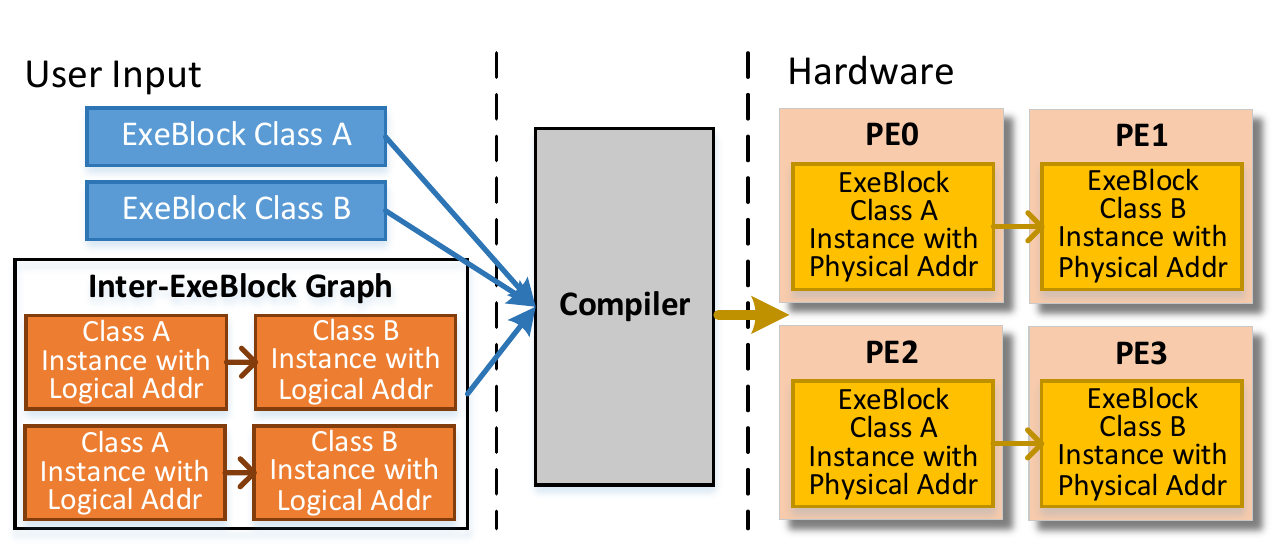}
\vspace{-0.05in}
\caption{Programming Model of \emph{RISC-NN}.} \label{fig:scheme:programming-model}
\vspace{-0.2in}
\end{figure} 

Figure \ref{fig:scheme:programming-model} shows the programming model for \emph{RISC-NN}:
\begin{compactitem}
\item \textbf{User Inputs}, which includes: 
\begin{compactitem}
\item \emph{Definition of ExeBlock Classes}. As will be discussed in Section \ref{sec:results201907:cnn}, a typical \emph{RISC-NN} task comprises of multiple \emph{ExeBlocks} with only few identical styles. A programmer needs to first provide the definition of the different \emph{ExeBlocks} styles (a.k.a., \emph{ExeBlock Classes}) which, just like \emph{ExeBlocks}, are composed of \emph{RISC-NN} instructions. Yet, their operation fields are address parameters instead of hardware physical addresses.
\item \emph{Execution Graph.} It consists of the instances of \emph{ExeBlocks Classes} and the graph defining the relationship of the \emph{ExeBlock Instances}. In \emph{ExeBlock Instances}, the address parameters are replaced by logical (hardware irrelevant) addresses. The programmer also needs to assign a logical PE id to each \emph{ExeBlock Instance}.

Note that, the data sharing is defined in \emph{Execution Graph}. For example, if the instructions of two \emph{ExeBlock Instances} will access the same operand data (e.g., \textbf{0x07} in \emph{ExeBlocks-0\&1} of Figure \ref{fig:scheme:data-reuse-sharing}), the programmer needs to first assign the same logical PE id to both \emph{ExeBlock Instances} to make sure the compiler will map them to the same physical PE. Then for the data-sharing instructions, the logical addresses of corresponding operands should be the same. 
\end{compactitem}

\item \textbf{Translator Responsibility.} We use a simple C++-based translator to handle the user inputs and maps them to the \emph{RISC-NN} hardware. Its responsibility includes: 
\begin{compactitem}
\item Mapping \emph{ExeBlocks} to physical PEs. The compiler optimizes the load-balancing among PE by considering the available resources of different PEs (i.e., current \emph{ExeBlock} count, available entries in {Instruction/Operand RAM Modules}, etc.).
\item Mapping logical in-PE addresses to physical entries in \emph{Operand RAM Module}. Here the compiler also balances the occupancy among the banks of each \emph{Operand RAM Module} to reduce the chance of \emph{Operand RAM} conflict (Section \ref{sec:scheme:control-bits}).
\item Mapping logical DRAM addresses to physical DRAM addresses.
\end{compactitem}
\end{compactitem}
\vspace{-10pt}

\ignore{
\subsection{Discussion: Complex Activation Functions.}\label{sec:scheme:discuss}

State-of-the-art \emph{NNs} need to use \emph{activation functions}, which convert the value of neurons into meaningful
outputs, between adjacent layers and before the final output. Conventional \emph{activation functions} (e.g., 
\emph{SoftMax}, \emph{Sigmoid}, \emph{TanH} and \emph{Gaussian}) are quite complex and usually require 
floating point operation. More recently, a set of simple \emph{activation functions} (e.g., \emph{LeRU},
\emph{Binary Step} and \emph{Identity}) were introduced. 

For design complexity and hardware cost considerations, we choose to keep the ISA of \emph{RISC-NN} simple and thus do not support complex \emph{activation functions}.
We argue that this results in little damage, if any, to the functionality of \emph{RISC-NN}, because recent researches
showed that, for most \emph{NN} applications, we can use simple \emph{activation functions} between intermediate 
layers without hurting the accuracy. For example, the recent \emph{TensorFlow} implementation of \emph{AlexNet} replaces 
\emph{SoftMax} with simple \emph{LeRU} between intermediate layers.  Though for most \emph{NN} applications, complex \emph{activation functions} 
are still needed for the final output, they require a very little fraction of overall computation. According to 
\emph{Amdale's Law}, there is little performance penalty if we execute these \emph{activation functions} in host 
machine, therefore there is no need to implement these \emph{activation functions} on \emph{RISC-NN} side.

In case complex \emph{activation functions} are a must for \emph{RISC-NN}, we can use prior art design 
solution \cite{DaDianNao}---each memory controller contains a light-weight \emph{Activation Table} which 
stores the low-precision results of the \emph{activation functions} for PEs to lookup.

}

\section{Experiment Methodology} \label{sec:method}
\begin{table}[t]
\caption{Detail of NN benchmarks  \label{table:methodology:convParameter}}
\centering
\footnotesize
\begin{tabular}{|l|l|l|l|l|} \hline
Layers/         & Batch  & Type &  Compress     \\ 
Operators       &  Size  &      &   Rate    \\ \hline
GoogLeNet\_CONV1 \cite{GoogLeNet}    & 128        & CNN   & N/A   \\ \hline
GoogLeNet\_CONV5a\_5 \cite{GoogLeNet} & 128        & CNN   & N/A   \\ \hline
VGG16\_CONV4 \cite{VGG}        & 64         & CNN   & 36\% \cite{2015deepcompression} \\ \hline
VGG16\_CONV9 \cite{VGG}        & 64         & CNN   & 27\% \cite{2015deepcompression} \\ \hline
VGG16\_CONV11 \cite{VGG}       & 64         & CNN   & 35\% \cite{2015deepcompression} \\ \hline
AlexNet\_CONV2 \cite{AlexNet}      & 256        & CNN   & 38\% \cite{2015deepcompression} \\ \hline
AlexNet\_CONV3 \cite{AlexNet}      & 256        & CNN   & 35\% \cite{2015deepcompression} \\ \hline
ResNet\_CONV2\_2 \cite{ResNet}    & 64         & CNN   & N/A   \\ \hline
ResNet\_CONV3\_1 \cite{ResNet}    & 64         & CNN   & N/A   \\ \hline
ResNet\_CONV4\_3 \cite{ResNet}    & 64         & CNN   & N/A   \\ \hline
Transformer\_ATT \cite{2017transformer}    & 128        & Attention  & N/A \\ \hline
Transformer\_MMM1 \cite{2017transformer}   & 128        & MMM   & N/A   \\ \hline
Transformer\_MMM2 \cite{2017transformer}   & 128        & MMM   & N/A   \\ \hline
Transformer\_FC1 \cite{2017transformer}    & 128        & MLP    & N/A   \\ \hline
Transformer\_FC2 \cite{2017transformer}    & 128        & MLP    & N/A   \\ \hline
Sentiment\_CNN1 \cite{2014Seq2CNN}     & 500        & 1D CNN & N/A   \\ \hline
Sentiment\_CNN2 \cite{2014Seq2CNN}     & 500        & 1D CNN & N/A   \\ \hline
Seq2seq\_LSTM \cite{Britz:2017}     & 32        & LSTM & N/A   \\ \hline
\end{tabular}
	  \vspace{-0.1in}
\end{table}

Table \ref{table:method:parameters} summarizes the parameters of a default \emph{RISC-NN} system. To evaluate the effectiveness of \emph{RISC-NN}, we use an in-house cycle-accurate simulator with Ramulator \cite{Ramulator} as its memory subsystem. We also implement the design in Verilog and synthesize it using Synopsys Design Compiler with  TSMC 12nm GP standard VT library. We then do circuit-level simulations of the atomic operations of \emph{RISC-NN} using Synopsys VCS and calculate energy consumption using PrimeTime PX. \revisioncolor{For host data transaction, we use the power consumption of PCIE 3.1\cite{Pciepower}}. Finally, we feed these energy data back into the simulator to estimate the overall energy consumption of the applications. 

As is shown in Table \ref{table:methodology:convParameter}, we measure the performance of \emph{RISC-NN} using layers/operators of multiple \emph{NN} inference applications, including classical \emph{CNN} models (GoogLeNet\cite{GoogLeNet}, AlexNet\cite{AlexNet} and VGG16\cite{VGG}) and other \emph{NN} models (ResNet50\cite{ResNet}, Transformer\cite{2017transformer}, a sentiment analysis model based \emph{CNN}\cite{2014Seq2CNN}) and seq2Seq\cite{Britz:2017} from MLPerf \cite{mlperf}. Table \ref{table:methodology:convParameter} shows the detail of these NN benchmarks. Table \ref{table:methodology:convParameter} also shows the compress rates of AlexNet and VGG16 layers \cite{2015deepcompression}, as we will use these layers as examples to show effectiveness of \emph{RISC-NN} for \emph{Sparse NN} applications. 

We compare energy consumption of \emph{RISC-NN} with state-of-the-art \emph{NVIDIA Titan Xp GPGPU} \cite{Titanxp}. To get accurate energy consumption of each \emph{NN} layer/operator, we use Caffe2 time \cite{Caffe} or Tensorflow timeline \cite{TensorFlow} to get execution time, and measure power consumption using Nvidia-smi \cite{nvidia-smi}. Finally, \emph{GPGPU} energy consumption is product of the measured execution time and power consumption.

\section{Experiment Results} \label{sec:results}
\subsection{Implementing ISAs of CISC NN Accelerators in RISC-NN}\label{sec:results201907:cisc}

\begin{table}[t]
	\caption{Supported and unnecessary \emph{CISC NN} accelerator instructions in the PE array of \emph{RISC-NN}.}\label{table:results201907:cisc-ins}
	\centering
	\footnotesize
		\setlength\tabcolsep{1.5pt} 
		\begin{tabular}{|l|l|l|} \hline
		               & TPU \cite{TPU}      & Cambricon \cite{Cambricon} \\ \hline
		Implementable    & Read\_Host\_Memory,      & VLOAD, VSTORE, MLOAD\\
		            & Read\_Weights,          & MSTORE, MMV, VMM\\
		               & MatrixMultiply,            & MMC, OP, MAM, MSM,  \\
		               & Convolve,                  & VAV, VAS, VMV, VSV, \\
		               & Activate (simple),         & VGTM, VGT, VE, VAND, \\
		               &                            & VOR,  VNOT, SC, SL \\ \hline
		Unnecessary    & Activate (complex)                           & jump, cond. branch, \\ 
		&          & VEXP, VDV, VLOG, RV \\ \hline
		\end{tabular}
	\vspace{-5pt}
\end{table}

\begin{figure*}[t]
\centering
\includegraphics[width=0.85\linewidth]{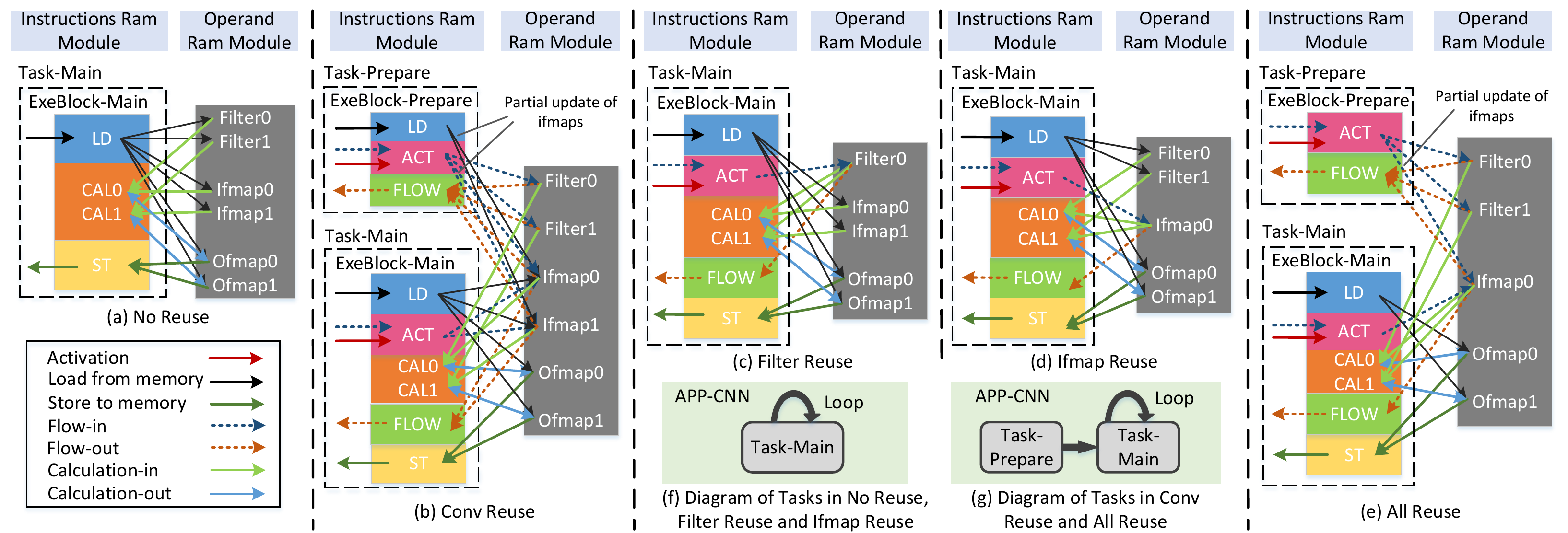}
	\vspace{-0.05in}
\caption{Dataflow diagram of Different CNN Implementations on \emph{RISC-NN}. Note that, to keep the graph simple, each \emph{ExeBlock} computes two sets of output feature map chunks. In the real implementation, each \emph{ExeBlock} computes four or eight sets of output feature map chunks.}
	\vspace{-0.1in}
\label{fig:results201907:convolution-implement}
\end{figure*}

\begin{table}[t]
\caption{Implementation Details of Instructions from CISC NN Accelerators on RISC-NN\label{table:results201907:ciscDetails}}
	\centering
	\footnotesize
\begin{tabular}{|l|l|l|l|l|l|l|l|}\hline
 & Size & LD & CAL & COPY & ST & Exe- & OP \\ 
 &      & Insts & Insts & Insts & Insts & Blocks   & RAM \\ 
 &		&			&			&		 &			&				& Entries \\ \hline
MMM  & 64x64 & 192  & 4096 & 4928 & 4096 & 255 & 5120 \\ \hline
MMV  & 64x64 & 4160 & 4096 & 525 & 64   & 255 & 8256 \\ \hline
MMS  & 64x64 & 4160 & 4096 & 0   & 4096 & 64  & 8256  \\ \hline
MAM  & 64x64 & 8192 & 4096 & 0   & 4096 & 64  & 12288 \\ \hline
OP   & 64x64 & 128  & 4096 & 896 & 4096 & 127 & 5120  \\ \hline
VGTM & 1024  & 2048 & 1024 & 0   & 1024 & 64  & 3072  \\ \hline
VMV  & 1024  & 2048 & 1024 & 0   & 1024 & 64  & 3072  \\ \hline
\end{tabular}
\vspace{-10pt}
\end{table}
\begin{table}[h]
\footnotesize
	\centering
	\vspace{-0.05in}
\caption{Static Program Analysis of Different CNN Implementations of AlexNet\_CONV2 on RISC-NN\label{table:results201907:CNN_Implementation_Details}}
\begin{tabular}{|l|l|l|l|l|l|l|l|}\hline
 & LD & CAL & COPY & ST & Exe- & OP & Max  \\ 
 & Insts & Insts & Insts & Insts & Blocks & RAM & Inst- \\ 
 &       &       &       &       &        & Entries & ances\\ \hline
No      & 13056 & 6400 & 0 & 256 & 64 & 13056 & 8 \\ 
Reuse & & & & & & &  \\ \hline
Conv    & 2976 & 6400 & 15200 & 256 & 256 & 13056 & 8\\
Reuse & & & & & & & \\ \hline
Filter  & 6681 & 6400 & 1575 & 256 & 120  & 8256 & 8 \\ 
Reuse & & & & & & & \\ \hline
Ifmap   & 6681 & 6400 & 1575 & 256 & 120 & 8256 & 8 \\
Reuse & & & & & & & \\ \hline
All    & 1136 & 6400 & 8400 & 256 & 254  & 8256 & 8 \\ 
Reuse & & & & & & & \\ \hline
\end{tabular}
	\vspace{-0.15in}
\end{table}

\begin{table}[t]
\footnotesize
\centering
\vspace{-0.03in}
\caption{ Best Performance's Instances number of Different CNN Implementations on RISC-NN\label{table:results201907:CNN_Instances_Details}}
\begin{tabular}{|l|l|l|l|l|l|}\hline
Layers & No    & Conv  & Filter & Ifmap  & All   \\ 
       & Reuse & Reuse & Reuse  & Reuse  & Reuse  \\ \hline
GoogLeNet\_CONV1     & 3 & 4 & 7 & 6 & 6  \\ \hline
GoogLeNet\_CONV- & 4 & 4 & 7 & 8 & 8  \\ 
-5a\_5               &   &   &   &   &    \\ \hline
VGG16\_CONV4         & 1 & 2 & 1 & 3 & 8  \\ \hline
VGG16\_CONV9         & 3 & 7 & 3 & 8 & 8  \\ \hline
VGG16\_CONV11        & 1 & 8 & 1 & 8 & 8  \\ \hline
AlexNet\_CONV2       & 3 & 6 & 4 & 8 & 8  \\ \hline
AlexNet\_CONV3       & 4 & 6 & 5 & 8 & 8  \\ \hline
ResNet\_CONV2\_2     & 1 & 3 & 2 & 6 & 8  \\ \hline
Average              & 2.5 & 5 & 3.75 & 6.88 & 7.75  \\ \hline
\end{tabular}
	\vspace{-5pt}
\end{table}

\begin{figure}[t]
    \centering
    \vspace{-0.15in}
    \includegraphics[width=\linewidth]{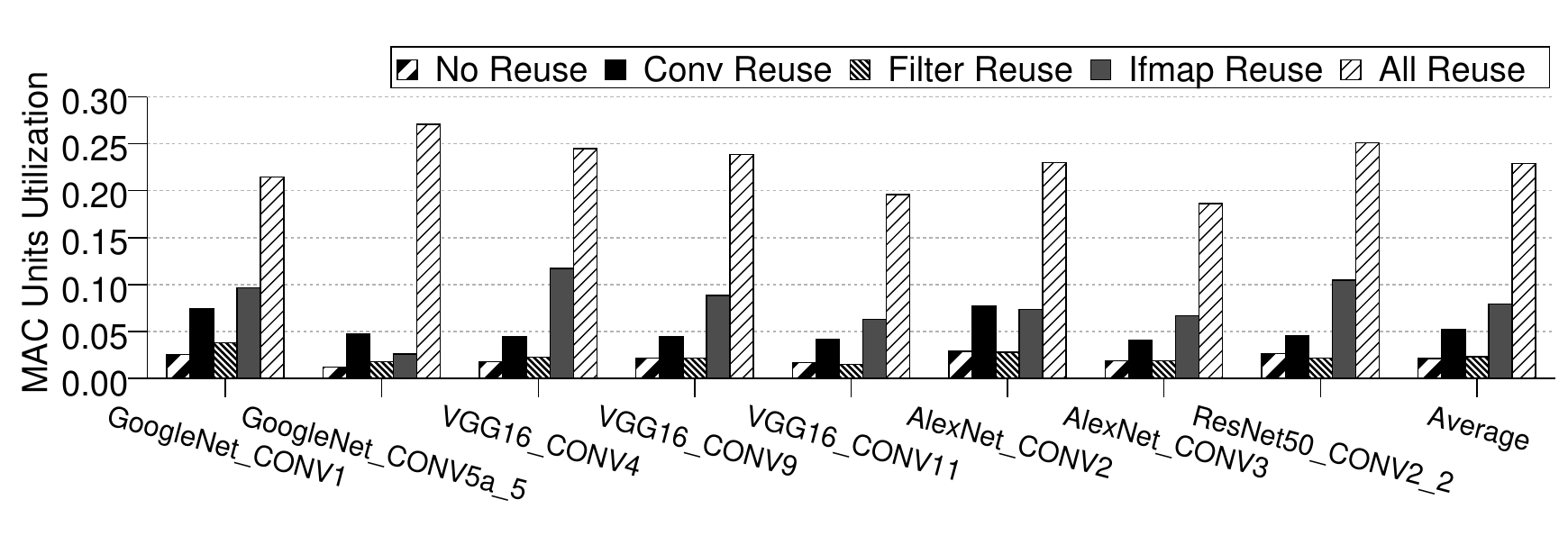}
    \vspace{-0.38in}
    \caption{MAC Units Utilization of Different CNN Implementations with Single Instances.}
    \label{fig:results201907:cnn-1task-comp-utility}    \centering
    \vspace{-0.05in}
    \includegraphics[width=\linewidth]{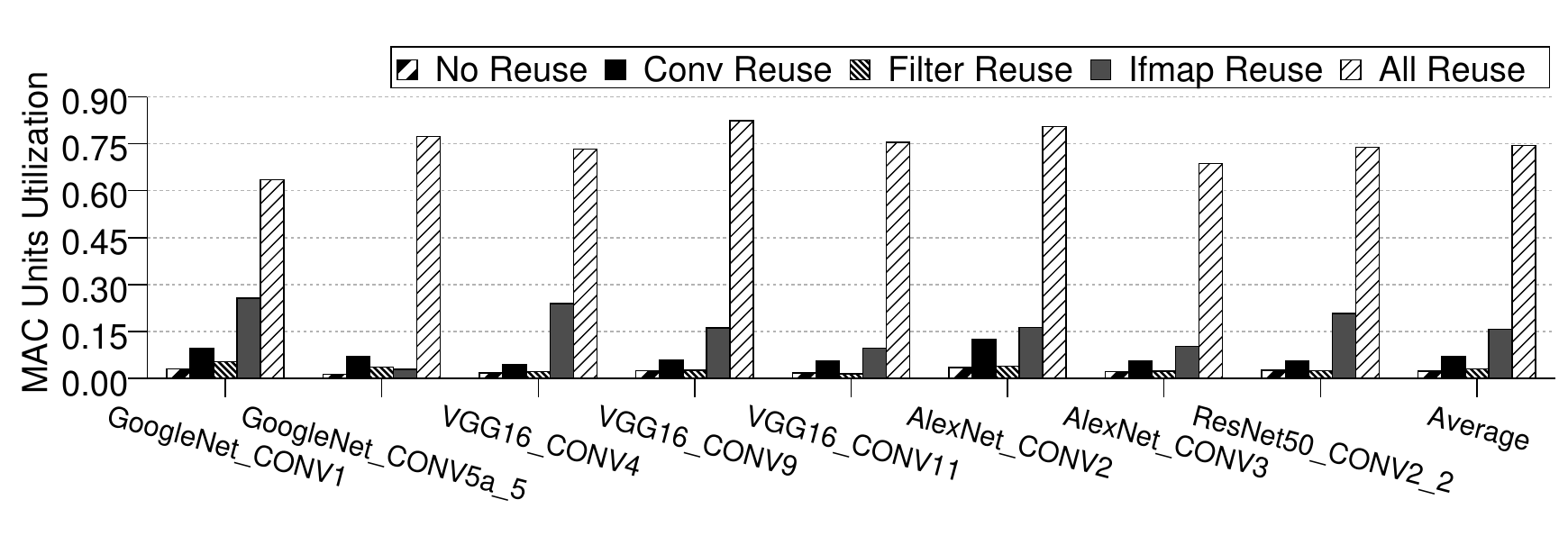}
    \vspace{-0.38in}
    \caption{MAC Units Best Utilization of Different CNN Implementations with Different Number of Instances.}
    \label{fig:results201907:cnn-best-comp-utility}    \centering
    \vspace{-0.25in}
\end{figure}

\begin{figure}[t]
    \centering
    \vspace{-10pt}
    \includegraphics[width=\linewidth]{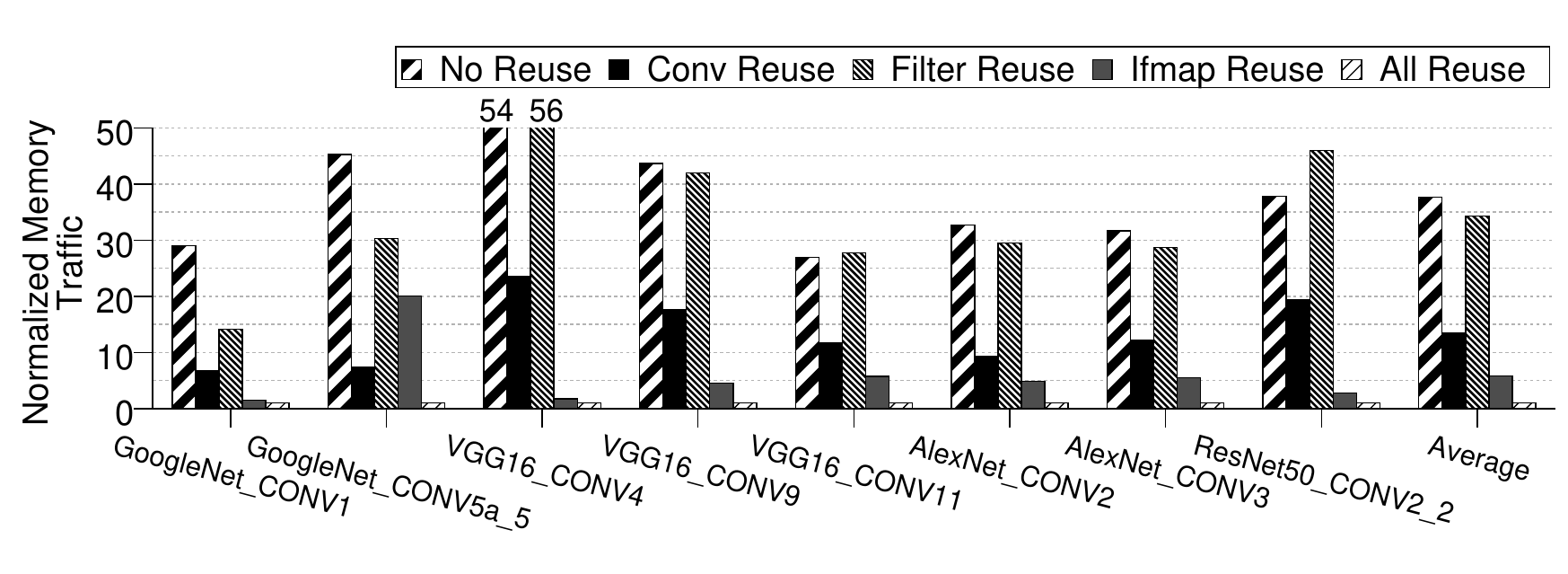}
    \vspace{-0.33in}
    \caption{Off-Chip Memory Traffic of Different CNN Implementations with Single Instance.}
    \label{fig:results201907:cnn-off-chip-traffic}    \centering
    \vspace{-0.20in}
\end{figure}
\begin{figure}[t]
   \centering
   \vspace{-0.05in}
    \includegraphics[width=\linewidth]{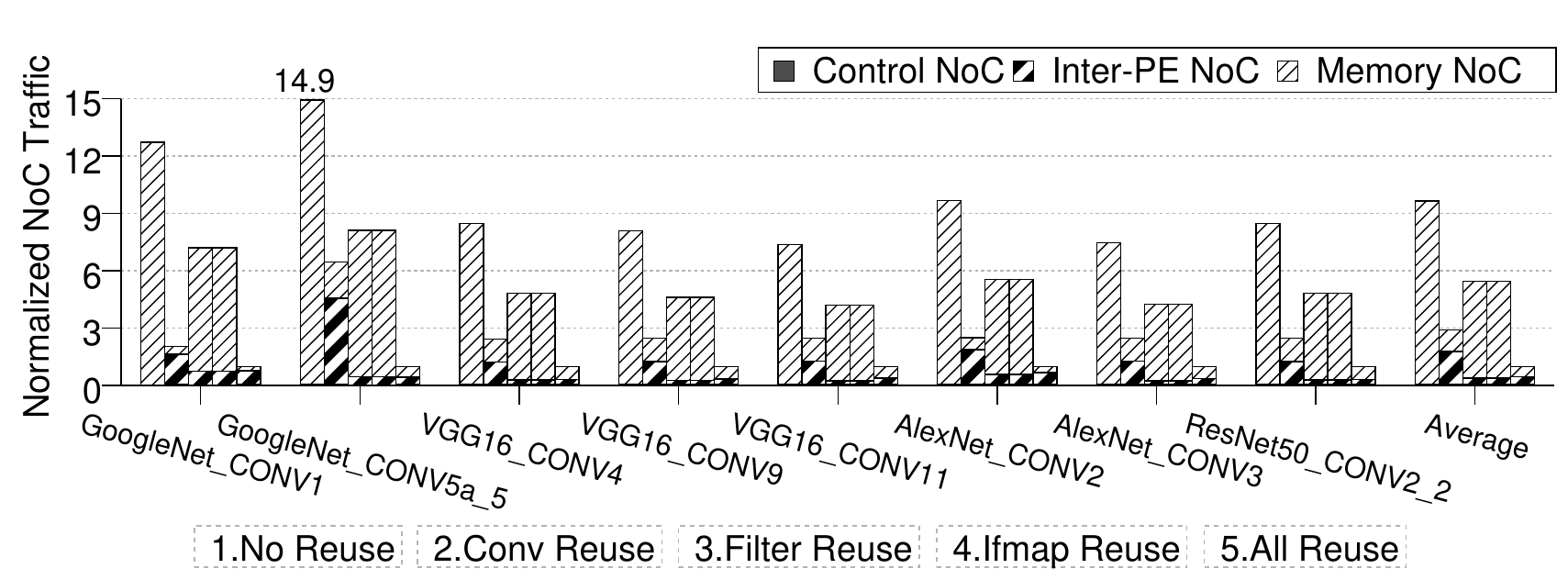}
    \vspace{-0.25in}
    \caption{NoC Traffic of Different CNN Implementations with Single Instance.}
    \label{fig:results201907:cnn-on-chip-traffic}    \centering
    \vspace{-0.23in}
\end{figure}
\begin{figure}[t]
    \centering
    \includegraphics[width=\linewidth]{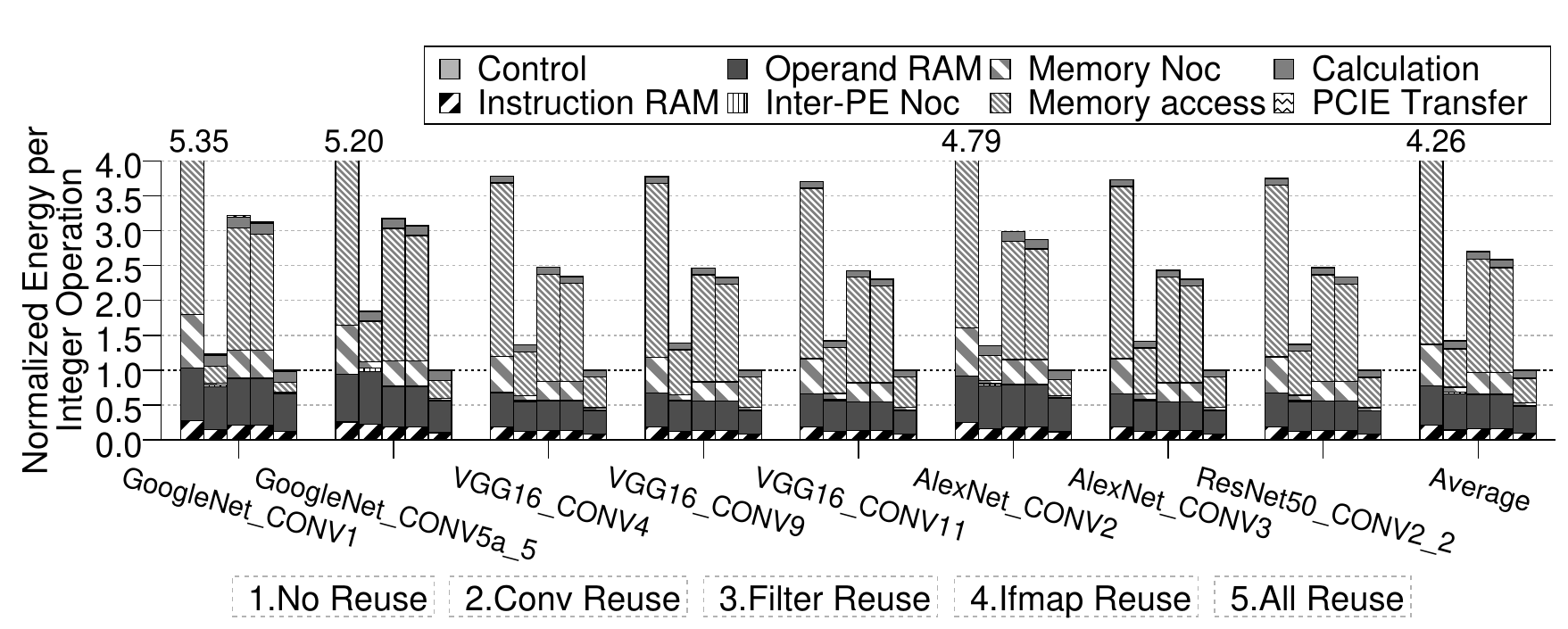}
    \vspace{-0.23in}
    \caption{Energy Efficiency of Different CNN Implementations.}
    \label{fig:results201907:cnn-energy}    \centering
    \vspace{-0.2in}
\end{figure}

\ignore{
\begin{figure*}[t]
	\vspace{-0.15in}
    \begin{minipage}[b]{0.24\linewidth}
        \centering
        \includegraphics[width=\linewidth]{pychart_drawing_new/Sparse-Oper-Ram.eps}
    \end{minipage}
    \begin{minipage}[b]{0.24\linewidth}
        \centering
        \includegraphics[width=\linewidth]{pychart_drawing_new/Sparse-Inst-Ram.eps}
    \end{minipage}
    \begin{minipage}[b]{0.24\linewidth}
        \centering
        \includegraphics[width=\linewidth]{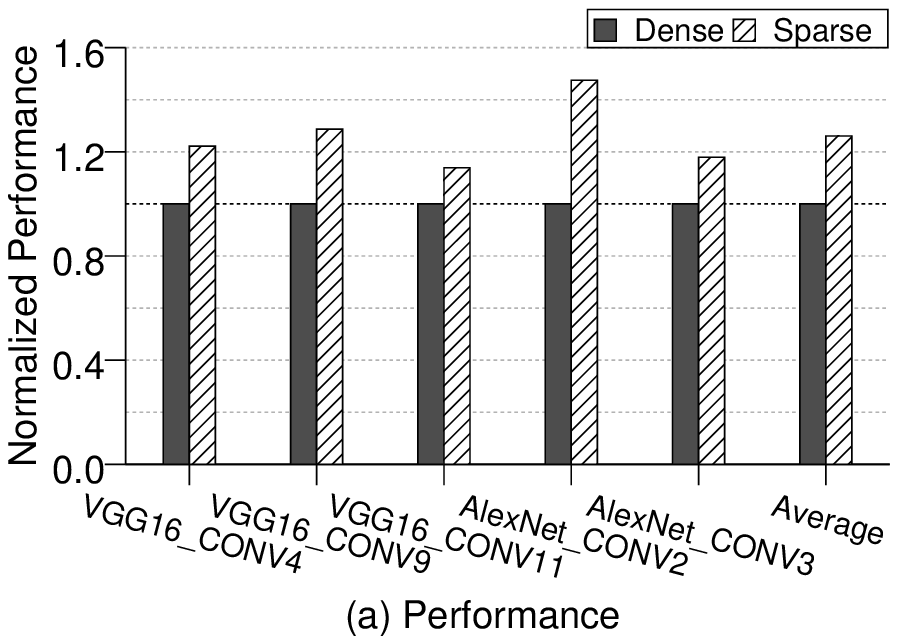}
    \end{minipage}
    \vspace{-0.05in}
    \begin{minipage}[b]{0.26\linewidth}
        \centering
        \includegraphics[width=\linewidth]{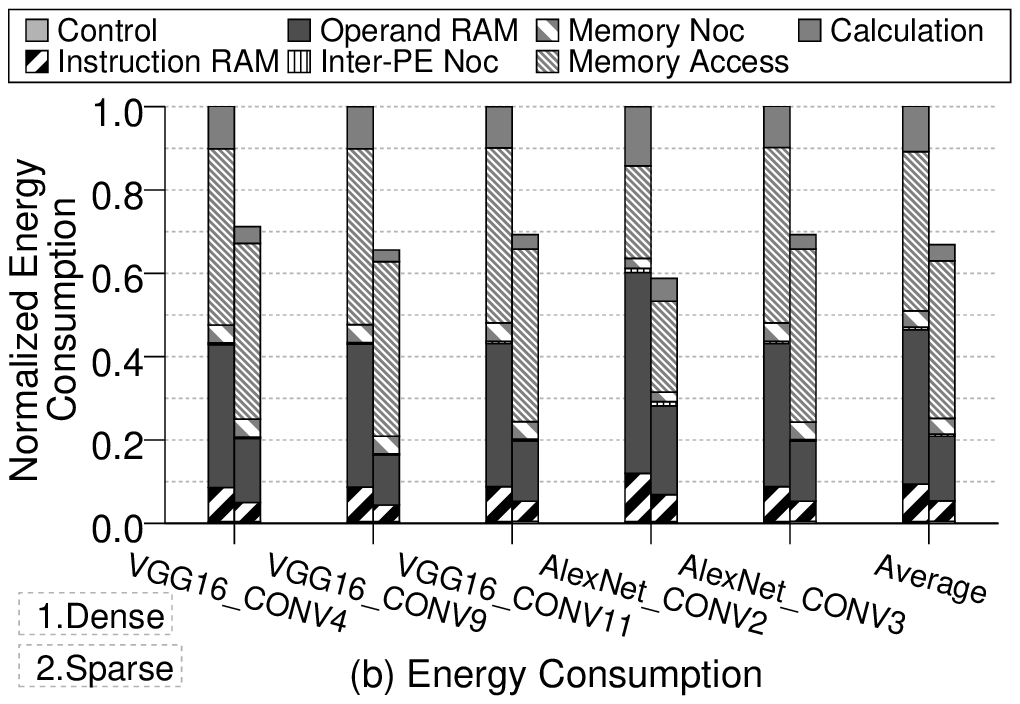}
    \end{minipage}
    \caption{Effectiveness of Sparse NN Optimization in RISC-NN}
    \label{fig:results201907:sparse-performance}
    \vspace{-0.1in}
    \centering
    \begin{minipage}[b]{0.65\linewidth}
        \centering
        \includegraphics[width=\linewidth]{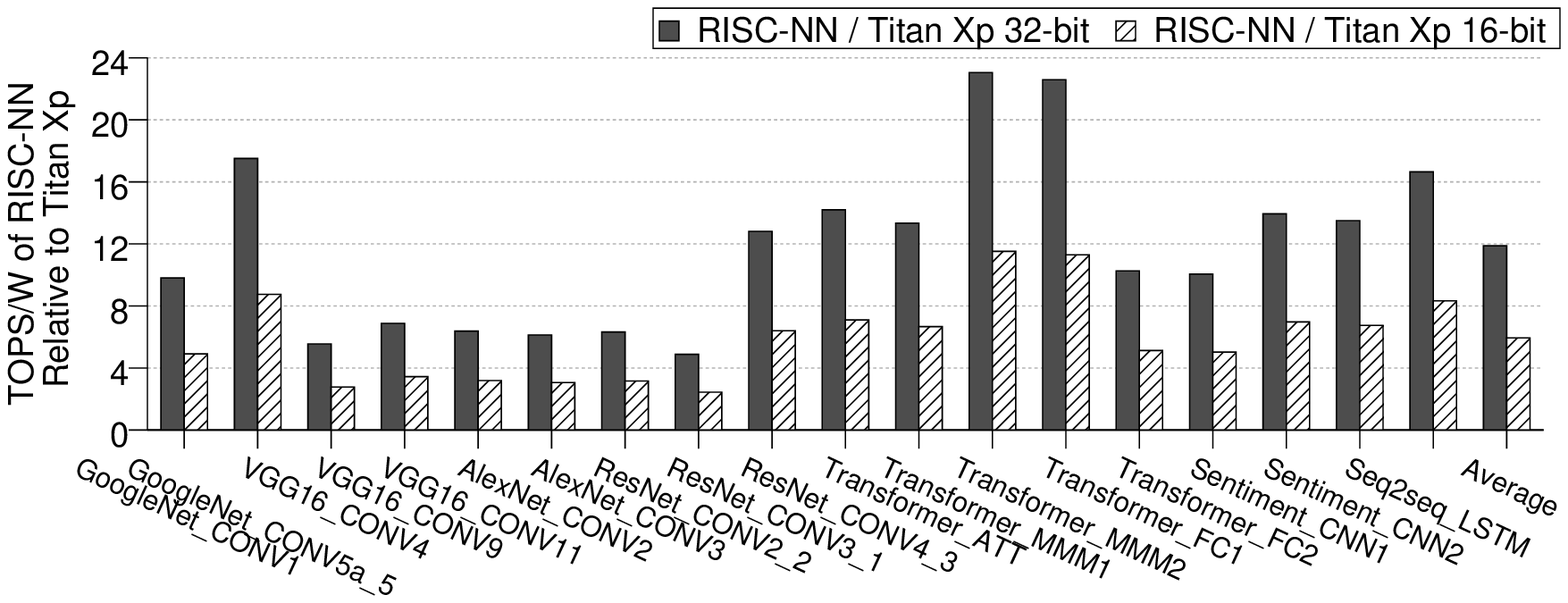}
    \end{minipage}
    \vspace{-0.05in}
    \begin{minipage}[b]{0.33\linewidth}
        \centering
        \includegraphics[width=\linewidth]{pychart_isca/Sparse_REDUCTION_ENERGY.eps}
    \end{minipage}
    \caption{Energy Reduction of \emph{RISC-NN} over \emph{Nvidia Titan Xp GPGPU}}
    \label{fig:results201907:energy-reduction}    \centering
\end{figure*}
}

\begin{figure}[t]
    \vspace{-0.12in}
    \centering
    \includegraphics[width=\linewidth]{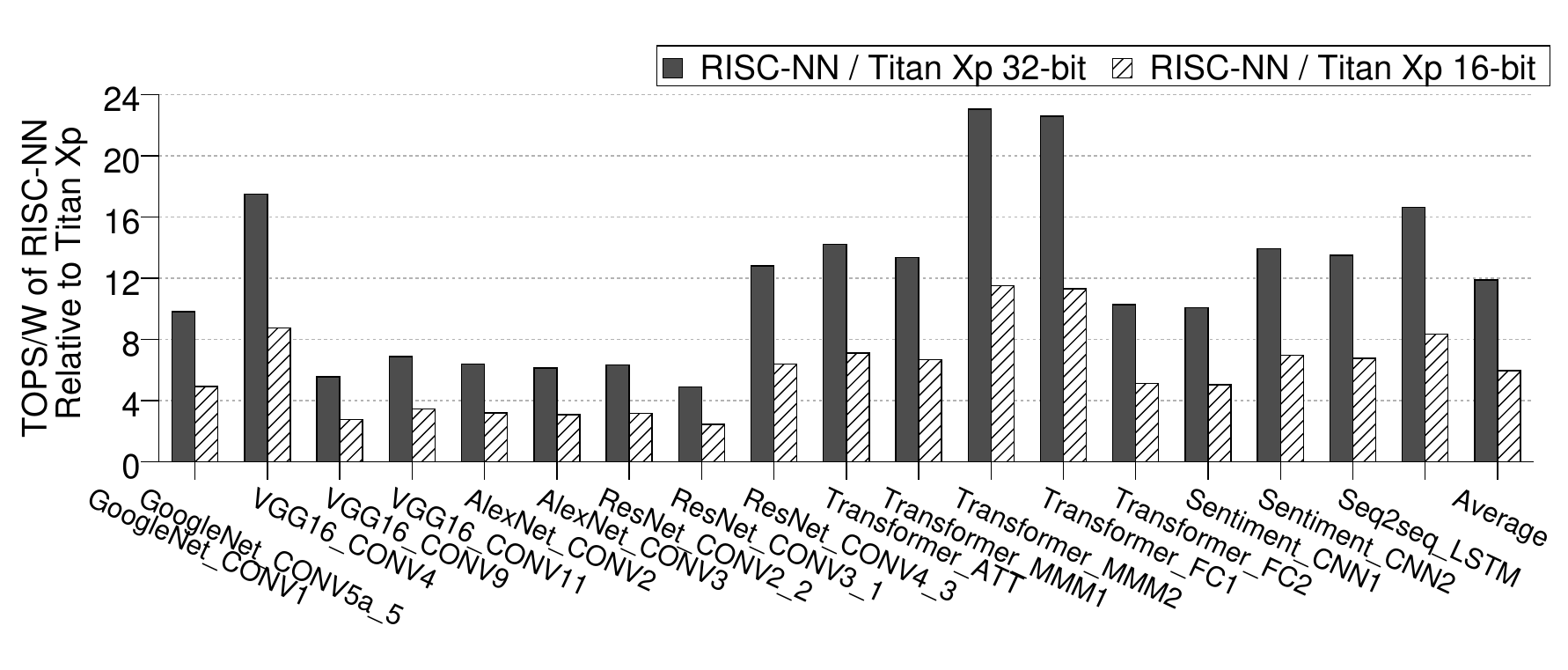}
    \vspace{-0.35in}
    \caption{Energy Reduction of \emph{RISC-NN} over \emph{Tian Xp GPGPU}}
    \label{fig:results201907:energy-reduction}    
    \centering
    \vspace{-0.25in}
\end{figure}

\begin{figure}[t]
    \vspace{-0.05in}
    \begin{minipage}[b]{0.52\linewidth}
        \centering
        \includegraphics[width=0.8\linewidth]{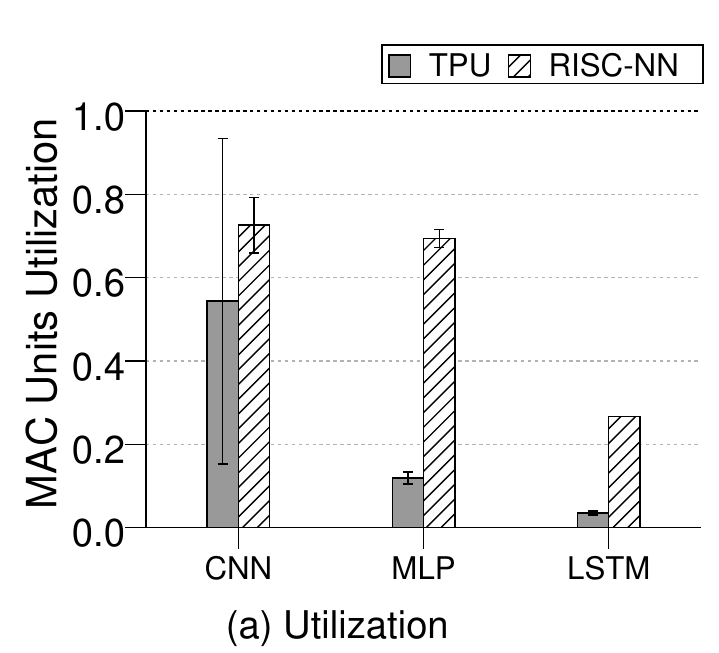}
    \end{minipage}
    \hspace{-0.3in}
    \vspace{-0.05in}
    \begin{minipage}[b]{0.48\linewidth}
        \centering
        \includegraphics[width=0.8\linewidth]{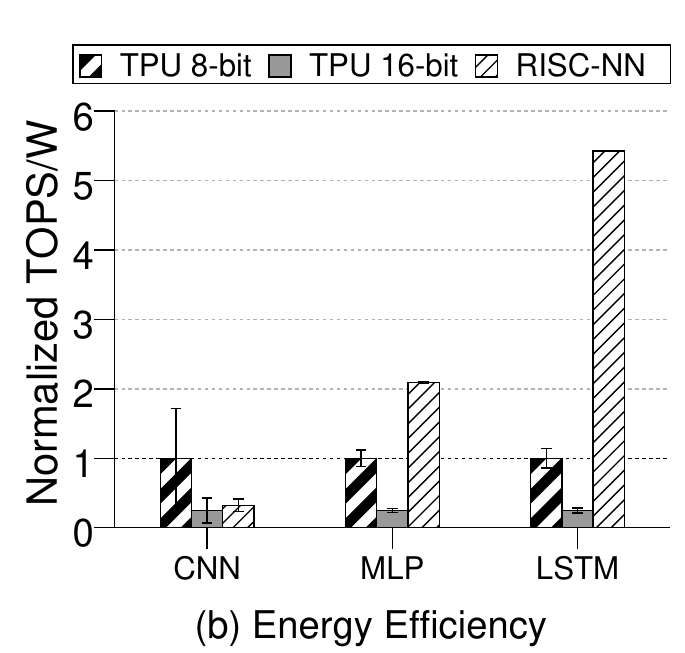}
    \end{minipage}
    \caption{Comparison of RISC-NN over TPU}
    \label{fig:results201907:tpu-comparison}
        \vspace{-0.23in}
\end{figure}

\begin{figure}[t]
\centering
    \vspace{0.07in}
    \includegraphics[width=\linewidth]{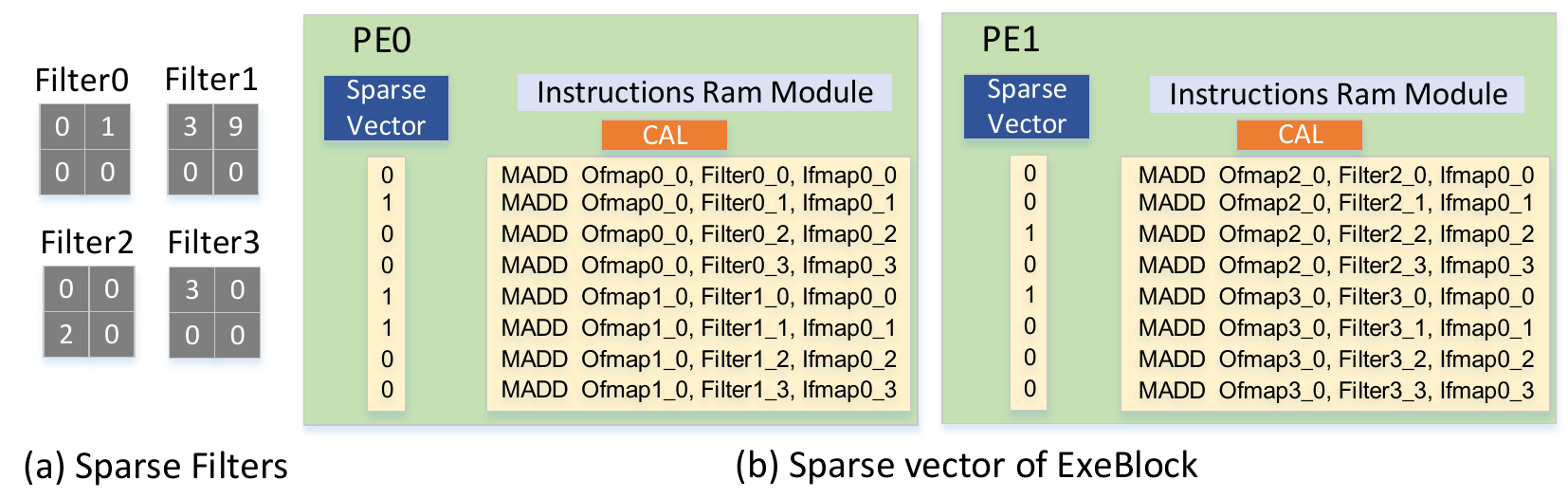}
    \vspace{-0.25in}
    \caption{Pruning Optimization for \emph{Sparse NN} in \emph{RISC-NN}. } \label{fig:sparse:idea}
\centering
    \begin{minipage}[b]{0.48\linewidth}
        \centering
        \includegraphics[width=\linewidth]{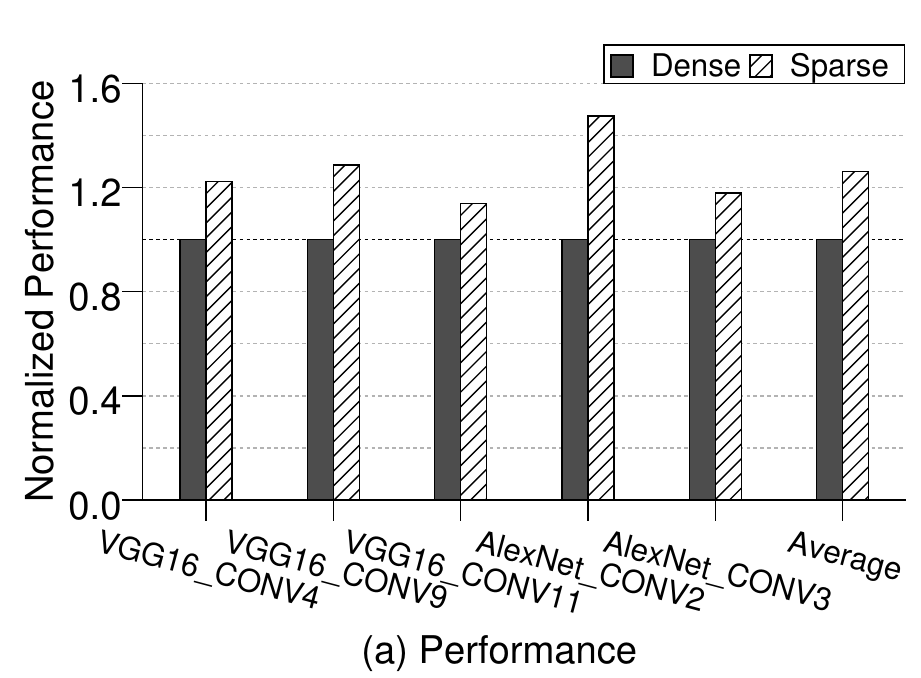}
    \end{minipage}
    \vspace{-0.05in}
    \begin{minipage}[b]{0.48\linewidth}
        \centering
        \includegraphics[width=\linewidth]{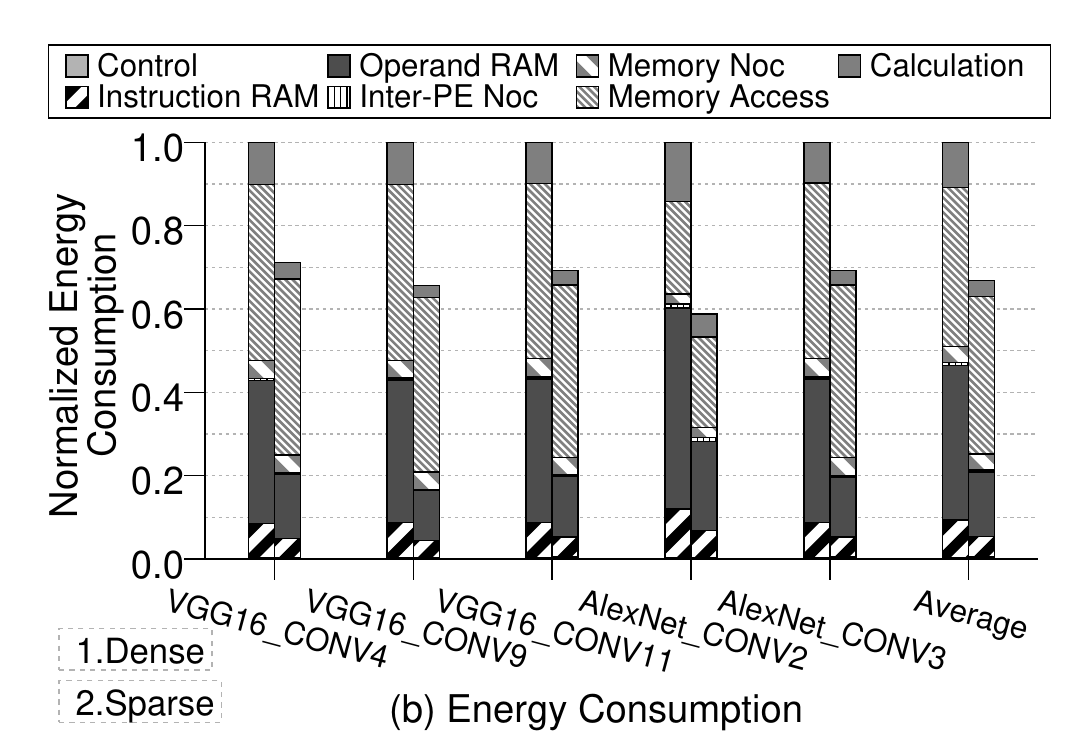}
    \end{minipage}
    \caption{Effectiveness of Sparse NN Optimization}
    \label{fig:results201907:sparse-performance}
\end{figure}

\begin{figure}[t]
    \vspace{-0.1in}
    \centering
    \begin{minipage}[b]{0.40\linewidth}
        \centering
        \includegraphics[width=\linewidth]{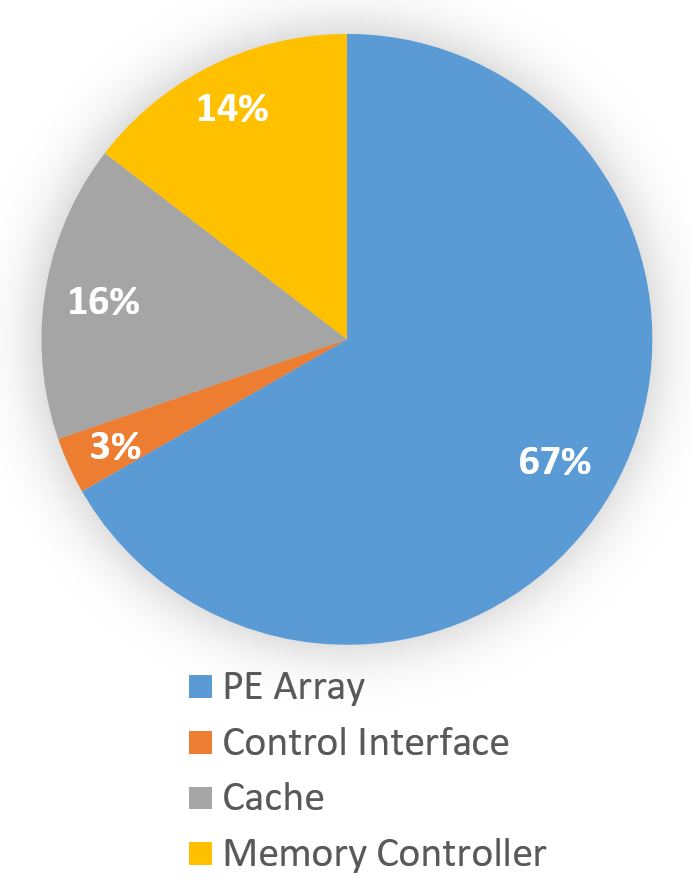}
        \vspace{-0.25in}
        \caption{Area Breakdown of RISC-NN}
        \label{fig:results201907:risc-area}
        \centering   
    \end{minipage}
    \quad
    \begin{minipage}[b]{0.53\linewidth}
        \centering
        \includegraphics[width=\linewidth]{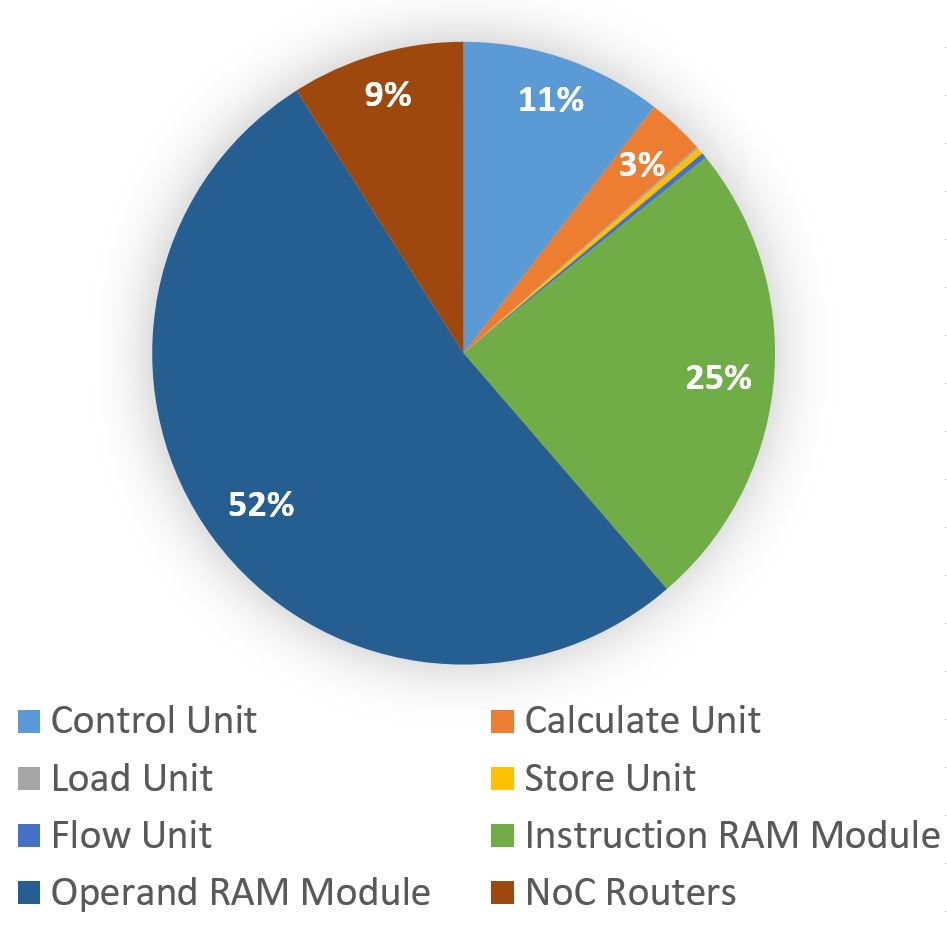}
        \vspace{-0.25in}
        \caption{Area Breakdown of a single RISC-NN PE}
        \label{fig:results201907:pe-area}  
        \centering
    \end{minipage}
    \vspace{-10pt}
\end{figure}

Table \ref{table:results201907:cisc-ins} summarizes the implementability of all the CISC instructions of TPU \cite{TPU} and Cambricon \cite{Cambricon} on \emph{RISC-NN}. We can see the following:
\begin{compactitem}
    \item \emph{RISC-NN} can implement most of these instructions.
    \item \emph{RISC-NN} can achieve the same behavior of two control instructions (i.e., jump and conditional branch) through inter-task management (either through host or control interface). Therefore these two instructions are not necessary for \emph{RISC-NN}.
    \item Five CISC instructions (complex activate, VEXP, VDV, VLOG and RV), which are mainly used for complex activation functions and classifiers, are unimplementable with \emph{RISC-NN} ISAs. Instead, as is discussed in Section \ref{sec:scheme:complex-funcs}, \emph{RISC-NN} uses an alternate solution which implements these functions as In-DRAM lookup tables.
\end{compactitem}

Table \ref{table:results201907:ciscDetails} shows the implementation details of several Cambricon \cite{Cambricon} CISC instructions on \emph{RISC-NN}. We can see that, all these implementations have quite a high ratio of memory instructions (LD and ST) to CAL instructions. As we will show in Section \ref{sec:results201907:cnn}, this phenomenon indicates that these instructions have poor data reuse within the PE array and thus are not well optimized.

\revisioncolor{
We acknowledge \emph{RISC-NN} has a higher code size overhead than \emph{CISC} accelerators \cite{Plasticine,StreamDataFlow,TPU,Cambricon}. However, as will be shown in the rest of the section, the benefits of \emph{RISC} clearly outweigh its code size overhead. Additionally, such overhead can be largely offset by two countermeasures: (1) When \emph{RISC-NN} executes the same application for multiple times (common case for cloud-based \emph{NN} accelerators \cite{TPU}), the instructions need to be loaded only once and reused thereafter; (2) Using SIMD also effectively offsets the overhead by sharing a single instruction on multiple data points.
}
\vspace{-15pt}
\subsection{Optimizing CNN in RISC-NN}\label{sec:results201907:cnn}

To show the flexibility and programmability of \emph{RISC-NN}, we analyze in detail of some typical CNN 2D convolution layers (see Table \ref{table:methodology:convParameter}). As is shown in Figure \ref{fig:results201907:convolution-implement}, following the category mechanism proposed by Chen et al. \cite{Eyeriss}, we evaluate five CNN implementations with different data reuse mechanisms on \emph{RISC-NN}. \emph{No Reuse} is the baseline scheme that has no data reuse, while \emph{Filter Reuse} and \emph{Ifmap Reuse} reuse the whole filters and input feature map chunks (both at the granularity of 2D filter size), respectively. These three implementations have only one task (i.e., \emph{Task-Main}), which loops itself multiple times. Meanwhile, \emph{Conv Reuse} introduces more complex partial inter-PE data sharing of the filters and input feature maps to further reduce off-chip memory traffic. To fulfill such data sharing, \emph{Conv Reuse} needs the assistance of a single-run preprocess task (i.e., \emph{Task-Prepare}). \emph{All Reuse} combines all the aforementioned reuse techniques together. Note that, \emph{Conv Reuse} and \emph{All Reuse}, which were implemented in a recent ASIC \emph{CNN} accelerator \cite{Eyeriss}, cannot be achieved in \emph{CISC-NN} accelerators \cite{Cambricon}\cite{TPU}. This fact indicates that \emph{RISC-NN} achieves higher programmability than \emph{CISC-NN} accelerators.

\ignore{\begin{table}[h]
\caption{Description of Parameters in Conv layers  \label{table:results201907:convParameter}}
\centering
\footnotesize
\begin{tabular}{|l|l|} \hline
Symbol & Description                            \\ \hline
H      & Width and height of an ifmap plane     \\ \hline
R      & Width and height of a filter plane     \\ \hline
C      & Number of ifmap/filter channels        \\ \hline
K      & Number of ofmap channels               \\ \hline
u      & Conv stride                            \\ \hline
\end{tabular}
\end{table}

\begin{table}[t]
\caption{Parameters of Evaluated Conv layers  \label{table:results201907:convlayers}}
\centering
\footnotesize
\begin{tabular}{|l|l|l|l|l|l|l|} \hline
Symbol              &  H  &  R  &  C  &  K  &  u  &  Description  \\ \hline
GoogleNet\_CONV1     & 224 &  7  &  3  &  64 &  2  &        \\ \hline
GoogleNet\_CONV5a\_5 &  7  &  5  &  32 & 128 &  1  &        \\ \hline
VGG16\_CONV4         & 112 &  3  & 128 & 128 &  1  &        \\ \hline
VGG16\_CONV9         &  28 &  3  & 512 & 512 &  1  &        \\ \hline
VGG16\_CONV11        &  14 &  3  & 512 & 512 &  1  &        \\ \hline
AlexNet\_CONV2       &  31 &  5  &  96 & 256 &  1  &        \\ \hline
AlexNet\_CONV3       &  15 &  3  & 256 & 384 &  1  &        \\ \hline
ResNet50\_CONV2\_2   &  56 &  3  &  64 &  64 &  1  &        \\ \hline
\end{tabular}
\end{table}
}

\subsubsection{Static Analysis of CNN Implementations.} 

Table \ref{table:results201907:CNN_Implementation_Details} shows the static program statistics of all the CNN implementations of AlexNet\_CONV2:
\ignore{Note that, since a 32$\times$32 input feature map is too large to be fit into \emph{Operand RAM Modules}, we divide it into 5$\times$5 chunks and calculate one chunk at a time in main calculation \emph{ExeBlocks} for all the CNN implementations. }
\begin{compactitem}
\item All implementations have the same number of CAL instructions.
\item \emph{No Reuse} has the highest number of \emph{LD} instructions because of its lack of data reuse. \emph{Conv Reuse}, \emph{Filter Reuse} and \emph{Ifmap Reuse} all substantially reduce the number of \emph{LD} instructions. \emph{All Reuse}, as the combination of all the reuse schemes, achieves the fewest \emph{LD} instructions and thus the lightest memory read traffic.
\item The implementations with reuse mechanisms all use \emph{COPY} instructions to reuse on-chip data among the \emph{ExeBlocks} (either intra-PE or inter-PE).
\item \emph{Filter Reuse}, \emph{Ifmap Reuse} and \emph{All Reuse} also reduce the number of \emph{Operand RAM} entries used by each task.
\end{compactitem}
As is shown in Table \ref{table:results201907:CNN_Implementation_Details}, \emph{RISC-NN} has enough \emph{Instruction} and \emph{Operand RAM Module} capacity to incorporate multiple instances of all the CNN implementation (8 in all five CNN implementations of AlexNet\_CONV2). Running multiple instances can increase \emph{ExeBlock}-level parallelism and thus improve the performance. We evaluation the performance impact of running multiple instances in Section \ref{sec:results201907:cnn:performance}.

\subsubsection{Performance of CNN Implementations.}\label{sec:results201907:cnn:performance}

Figure \ref{fig:results201907:cnn-1task-comp-utility} shows the performance of \emph{RISC-NN} in terms of MAC utilization under different CNN implementations with single instance---\emph{All Reuse} achieves an average MAC unit utilization of 22.91\%, remarkably surpassing the results of others (2.12\%, 5.21\%, 2.30\% and 7.96\% respectively for \emph{No Reuse}, \emph{Conv Reuse}, \emph{Filter Reuse} and \emph{Ifmap Reuse}). 

Due to conflicts of shared resources, such as NOCs and offchip-Memory, having more instances does not necessarily lead to higher performance. Table \ref{table:results201907:CNN_Implementation_Details} shows the instances number of the best MAC unit utilization for five different CNN implementations. The performance of \emph{All Reuse} keeps increasing as its number of instances increases to about 8; however, this is not the case for other implementations---\emph{No Reuse}, \emph{Conv Reuse}, \emph{Filter Reuse} and \emph{Ifmap Reuse} achieve their peak performance with an instance number of on average 2.5, 5, 3.75 and 6.88, respectively, and adopting more instances beyond these sweet spots hurts the performance. As is shown in Figure \ref{fig:results201907:cnn-best-comp-utility}, for the multi-instance situation, the best MAC unit utilization of \emph{All Reuse} (on average 74.43\%) is at least more than 2 times higher than the other implementations.

\ignore{
 Our analysis shows that this seemingly counterintuitive phenomenon is mainly caused by the following conflicts:
\begin{compactitem}
    \item \emph{Conflicts of Off-Chip Memory.} The current off-chip memory subsystem of \emph{RISC-NN} is not conflict-aware. Therefore, the conflict overhead might be high when there is an excessive amount of memory traffic generated by multiple program instances. For \emph{All Reuse}, this phenomenon is mild because it is heavily optimized to reduce off-chip memory traffic, as we show in Section \ref{sec:results201907:cnn:traffic}. This fact also indicates that, as with the general purpose RISC processors, \emph{RISC-NN} relies heavily on software optimization.
   \item \emph{Conflicts of Operand RAM Modules.} As shown in Section \ref{sec:scheme:control-bits}, when there is an \emph{Operand RAM Module} conflict for a \emph{CAL} instruction, we insert \emph{NOP} instruction(s) to solve it, and this hurts the performance. \emph{Operand RAM Module} conflicts are the major reason why the performance of \emph{All Reuse} becomes relatively saturated with more instances---e.g., for the 7-instance \emph{All Reuse}, 22\% of the execution time is wasted on \emph{NOPs}. Part of the reason for this is that we currently semi-manually allocate \emph{Operand Module RAM} entries, which is far from ideal. We believe a cleverer \emph{Operand Module RAM} allocation policy can largely solve this issue, and this will be our future work.
\end{compactitem}
}

\ignore{
\begin{table*}[t]
\footnotesize
	\centering
\caption{ Best Performance's Instances number of Different CNN Implementations on RISC-NN\label{table:results201907:CNN_Instances_Details}}
\begin{tabular}{|l|l|l|l|l|l|l|l|l|l|}\hline
 & GoogleNet & GoogleNet & VGG16 & VGG16 & VGG16 & AlexNet & AlexNet & ResNet  \\ 
 & \_CONV1   & \_CONV5a\_5 & \_CONV4 & \_CONV9 & \_CONV11 & \_CONV2 & \_CONV3 & \_CONV2\_2 \\ \hline
No Reuse      & 3 & 4 & 1 & 7 & 3 & 3 & 5 & 2 \\ \hline
Conv Reuse    & 2 & 4 & 1 & 7 & 7 & 6 & 5 & 3 \\ \hline
Filter Reuse  & 5 & 2 & 1 & 7 & 3 & 4 & 5 & 2 \\ \hline
Ifmap Reuse   & 5 & 8 & 3 & 8 & 8 & 8 & 8 & 6 \\ \hline
All Reuse     & 7 & 8 & 8 & 8 & 8 & 8 & 8 & 8 \\ \hline
\end{tabular}
\end{table*}
}

\subsubsection{Memory and NoC Traffic of CNN Implementations.}\label{sec:results201907:cnn:traffic}

Figures \ref{fig:results201907:cnn-off-chip-traffic} and \ref{fig:results201907:cnn-on-chip-traffic} present the memory and NoC traffic of different CNN implementations with single instance. We can see that, as expected, \emph{All Reuse} has the least memory traffic---only on average $\frac{1}{38}$, $\frac{1}{13}$, $\frac{1}{34}$ and $\frac{1}{6}$ of \emph{No Reuse}, \emph{Conv Reuse}, \emph{Filter Reuse} and \emph{Ifmap Reuse}, respectively. Note that, \emph{Filter Reuse} has considerably more memory accesses than \emph{Ifmap Reuse} because the filters of a CNN application usually have a smaller memory footprint than its input feature maps. This is also the reason why DaDianNao \cite{DaDianNao} and TPU \cite{TPU} use large on-chip memories to make input feature maps stay on-chip. \emph{Conv Reuse} has less memory traffic than \emph{Filter Reuse} because it also partially reduces the access to off-chip feature maps.

On the NoC traffic side, \emph{Memory NoC} traffic is proportional to the \emph{Off-Chip Memory} traffic shown in Figure \ref{fig:results201907:cnn-off-chip-traffic}, and the \emph{Control NoC} traffic takes a quite small portion of all NoC traffic (less than 8\% for all implementations). While \emph{No Reuse} has no \emph{Inter-PE NoC} traffic since it has no inter-PE data sharing, all the other implementations have a significant amount of \emph{Inter-PE} traffic to realize their unique inter-PE data-sharing pattern using \emph{COPY} instructions. An interesting phenomenon is that there is a mismatch of off-chip memory traffic and memory NoC traffic in \emph{Ifmap Reuse}---compared to \emph{All Reuse}, \emph{Ifmap Reuse} has nearly on average 9.39$\times$ more \emph{Memory NoC} traffic but only on average 5.82$\times$ more off-chip memory traffic. This happens because the memory request of \emph{Ifmap Reuse}, the majority of which is input feature map information, achieves a high hit rate (at least 91.92\%) in the front-end cache of the memory controller.

\subsubsection{Energy Efficiency of CNN Implementations.}\label{sec:results201907:cnn:energy}

Figure \ref{fig:results201907:cnn-energy} shows the normalized energy consumption of all the CNN implementations. We can see that, consistent with NoC and off-chip memory traffic statistics, \emph{No Reuse} consumes the highest energy while \emph{All Reuse} consumes the lowest energy.

\subsection{Performance Efficiency Over State-of-the-Art GPGPU and CISC NN Accelerator}\label{sec:results201907:gpgpu}

We compare performance efficiency (in terms of TOPS/W) of \emph{RISC-NN} with state-of-the-art \emph{Nvidia Titan Xp} GPGPU and TPU \cite{TPU}. 

\textbf{RISC-NN VS. GPGPU.} Since \emph{Titan} uses 32-bit accuracy instead of 16-bit (the accuracy of \emph{RISC-NN} implementation), to ensure fair comparison, we also provide extrapolated performance efficiency results of the GPGPU with 16-bit accuracy. As a stream processor, the major part of GPGPU energy is consumed during data movement and such energy is proportional to the number of data moved, therefore we extrapolate that 16-bit accuracy GPGPU has roughly half the energy consumption of its 32-bit counterpart. We can see that, \emph{RISC-NN} achieves on average 11.88$\times$ and 5.94$\times$ higher energy performance efficiency over \emph{Nvidia Titan Xp} with 32-bit and 16-bit accuracy, respectively.

\textbf{RISC-NN VS. TPU.} There were limited performance/power details provided in \emph{CISC-NN} processors (i.e., \emph{Cambricon} and \emph{TPU}). As the best efforts, we compare the average reported MAC unit utilization and performance efficiency of different types of \emph{NN} applications (i.e., \emph{CNN} and \emph{MLP}) in \emph{RISC-NN} and \emph{TPU} \cite{TPU}. As is shown in Figure \ref{fig:results201907:tpu-comparison}(a), \emph{RISC-NN} manages to achieve 72.62\%, 69.44\% and 26.66\% MAC unit utilization for \emph{CNN}, \emph{MLP} and \emph{LSTM}; while TPU achieves respectively 54.40\% (with high variability), 11.96\% and only 3.53\% for \emph{CNN}, \emph{MLP} and \emph{LSTM}. This indicates that, compared with \emph{CISC NN accelerators}, \emph{RISC-NN} manages to better utilize its hardware resources. As a result of better hardware utilization, \emph{RISC-NN} also achieves better performance efficiency than \emph{TPU} with the same 16-bit accuracy---as is shown in Figure \ref{fig:results201907:tpu-comparison}(b), \emph{RISC-NN} achieves on average 1.29$\times$, 8.37$\times$ and 21.71$\times$ performance efficiency over TPU for \emph{CNN}, \emph{MLP} and \emph{LSTM}, respectively. 


\vspace{-5pt}
\subsection{Sparse NN Optimization}\label{sec:results201907:sparse}

Here we use five \emph{Sparse CNN} layers \cite{2015deepcompression} as an example to show how \emph{RISC-NN} takes advantage of the sparsity. As is shown in Figure \ref{fig:sparse:idea}, the compiler first identifies ineffective weights of filters and generates corresponding \emph{sparse vector} to record the unnecessary instructions in next task. Then, as is discussed in Section \ref{sec:scheme:exeblock-manage}, \emph{Instruction Loader} takes the \emph{sparse vector} and translates it into \emph{Sparse PC Inc} in \emph{Instruction RAM}. Finally, \emph{CAL Unit} skips unnecessary instructions according to information of \emph{Sparse PC Inc} in each instruction.

As is shown in Figure \ref{fig:results201907:sparse-performance}, this approach boosts the performance by on average 26.06\%, as a result of skipping unnecessary instructions. It also reduces energy consumption by on average 33.13\% because of reduction in calculation and accesses to \emph{Instruction} and \emph{Operand RAMs}.

\vspace{-5pt}
\subsection{Area Overhead of RISC-NN}\label{sec:results201907:hardware}

The total area of \emph{RISC-NN} is 14.42$mm^2$, and a single \emph{RISC-NN} PE takes 0.15$mm^2$.
Figure \ref{fig:results201907:risc-area} and \ref{fig:results201907:pe-area} show the area breakdowns of the whole \emph{RISC-NN} and a single PE, respectively. We can see that, PE array consumes the majority (67\%) of the whole chip area while cache takes only 16\%; in the meantime, the area of a single PE is dominated by \emph{Operand RAM Module}. Such an architecture with small shared cache and large in-PE memory is in line with our design principle of maximizing the data sharing within PEs and reducing transmission overhead of accessing remote data (see Section \ref{sec:scheme:data-reuse}).

\ignore{
\begin{figure}[h]
    \centering
    \includegraphics[width=0.63\linewidth]{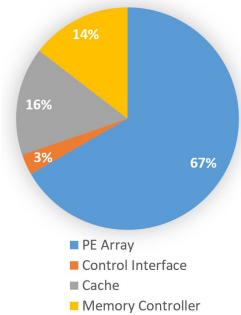}
    \caption{Area Breakdown of RISC-NN.}
    \label{fig:results201907:risc-area}    \centering
\end{figure}

\begin{figure}[h]
    \centering
 
    \includegraphics[width=0.65\linewidth]{pychart_drawing_new/PE-AREA.jpg}
    \caption{Area Breakdown of a RISC-NN PE.}
    \label{fig:results201907:pe-area}    \centering
\end{figure}
}

\ignore{
Table \ref{table:results201907:AreaBreakdown} shows the area of a PE and its max frequency with different SIMD counts, and the area percentage breakdown with different SIMD counts. We can see that, we can effectively bring down the area cost of the control hardware of \emph{RISC-NN} by increasing the SIMD count---when we use SIMD-64, \emph{Control Unit} and \emph{Instruction RAM Module} only take 0.81\% and 2.74\% of total PE area, respectively. Also, using an aggressive SIMD structure has relatively low impact on the system frequency.

\begin{table*}[h]
\caption{Area Breakdown of a RISC-NN PE \label{table:results201907:AreaBreakdown}}
\centering
\footnotesize
\begin{tabular}{|l|l|l|l|l|l|l|l|}\hline
Unit                    & SIMD1   & SIMD2   & SIMD4   & SIMD8   & SIMD16  & SIMD32  & SIMD64  \\ \hline
Total Area ($mm^2$)     & 0.055   & 0.065   & 0.086   & 0.128   & 0.214   & 0.394   & 0.754   \\ \hline
Max Frequency (Mhz)     & 1887    & 1887    & 1887    & 1887    & 1887    & 1887    & 1818    \\ \hline
Control Unit            & 11.14\% & 9.34\%  & 7.07\%  & 4.77\%  & 2.85\%  & 1.55\%  & 0.81\%  \\ \hline
Calculation Unit        & 1.28\%  & 1.84\%  & 2.53\%  & 3.08\%  & 3.63\%  & 3.63\%  & 3.56\%  \\ \hline
Load Unit               & 0.35\%  & 0.29\%  & 0.22\%  & 0.15\%  & 0.09\%  & 0.05\%  & 0.03\%  \\ \hline
Store Unit              & 0.49\%  & 0.48\%  & 0.48\%  & 0.47\%  & 0.47\%  & 0.45\%  & 0.42\%  \\ \hline
Flow Unit               & 0.32\%  & 0.34\%  & 0.37\%  & 0.40\%  & 0.46\%  & 0.44\%  & 0.43\%  \\ \hline
Instruction RAM Module  & 37.81\% & 31.68\% & 23.99\% & 16.40\% & 9.67\%  & 5.25\%  & 2.74\%  \\ \hline
Operand RAM Module      & 37.60\% & 44.39\% & 52.93\% & 61.74\% & 69.15\% & 74.93\% & 78.27\% \\ \hline
NoC Routers             & 11.02\% & 11.60\% & 12.38\% & 13.18\% & 13.69\% & 13.70\% & 13.74\% \\ \hline
\end{tabular}
\end{table*}
}
\vspace{-5pt}
\revisioncolor{
\subsection{Impact of SIMD on Energy Efficiency}\label{sec:results201907:simd}
\begin{figure}[h]
    \centering
    \vspace{-15pt}
    \includegraphics[width=\linewidth]{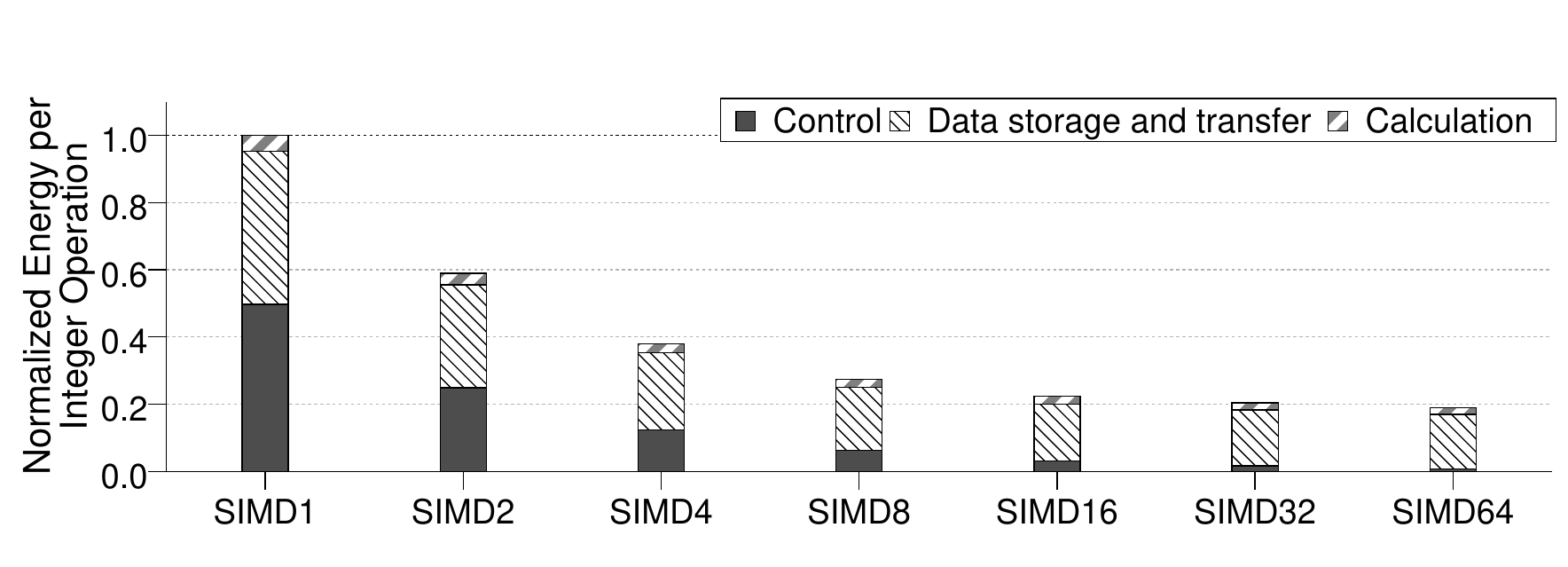}
    \vspace{-0.3in}
    \caption{Energy Efficiency of \emph{All Reuse} \texttt{AlexNet\_CONV2} with Different SIMD Width.}
    \label{fig:results201907:simd-efficiency}
\end{figure}\vspace{-5pt}
Using a high SIMD count in \emph{RISC-NN} not only reduces the area overhead of the control hardware, but also increase energy efficiency. Figure \ref{fig:results201907:simd-efficiency} shows the normalized energy efficiency (in terms of nJ/op) of the \emph{All Reuse} implementation of \texttt{AlexNet\_CONV2} under different SIMD counts. We can conclude that, the energy overhead of the control hardware diminishes as the SIMD count increases. With SIMD-64, the control energy only takes 0.8\% of the total energy consumption. Overall, we can achieve a reasonable control energy overall with SIMD-8.
}
\vspace{-5pt}
\revisioncolor{
\subsection{Energy Efficiency Scalability}\label{sec:reNNsults201907:scalability}
\begin{figure}[h]
    \centering
    \vspace{-15pt}
    \includegraphics[width=\linewidth]{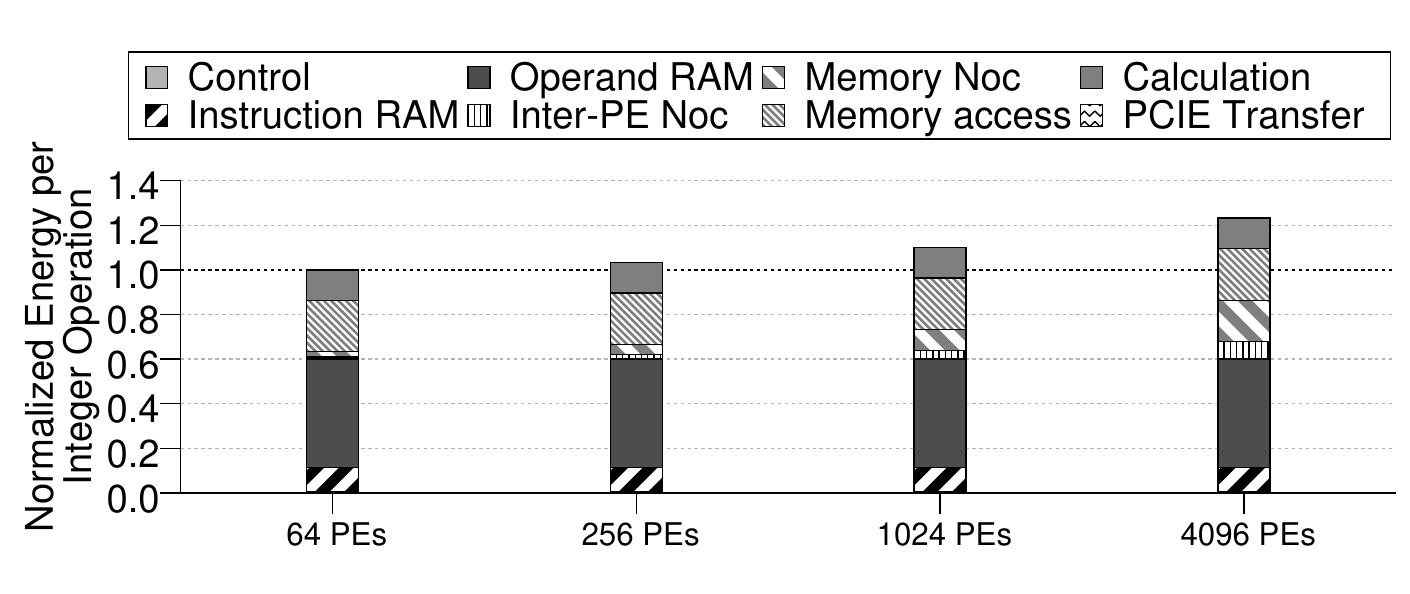}
    \vspace{-0.3in}
    \caption{Energy Efficiency Projection of \emph{All Reuse} \texttt{AlexNet\_CONV2} with Different Number of PEs.}
    \label{fig:results201907:scalability}
\end{figure}\vspace{-5pt}

As the number of PE increases in \emph{RISC-NN}, the energy efficiency of its components should stay largely the same, except for the NOC which on average requires more hops to finish a single request. Without loss of generality, we assume the number of hops and energy consumption per NOC request increase as the square root of PE counts. We then project the energy efficiency of \emph{RISC-NN} at different scales. As is shown in Figure \ref{fig:results201907:scalability}, when running \emph{All Reuse} applications, the energy efficiency of \emph{RISC-NN} scales well---even with the extreme scale of 4096 PEs, it energy efficiency is still just 23.1\% higher than that of the default 64-PE version, thanks to the optimizations of \emph{All Reuse} which keeps the data movement minimal.
}

\section{Discussion} \label{sec:discuss}
\begin{table*}[h]
\caption{Area Breakdown of RISC-NN and IoT RISC-NN \label{table:discuss:area-compare}}
\centering
\footnotesize
\begin{tabular}{|l|l|l|l|l|}\hline
\multirow{2}{*}{PE Units}          & \multicolumn{2}{|c|}{RISC-NN} & \multicolumn{2}{|c|}{IoT RISC-NN}     \\ \cline{2-5}
                        & config   & area ($mm^2$) & config   & area ($mm^2$) \\ \hline
Control Unit            & 32 Exeblocks; Unit-Level Concurrency & $1.58 \times 10^{-2}$ & 3 Exeblocks; No Unit-Level Concurrency & $1.12 \times 10^{-3}$    \\ \hline
Calculation Unit        & SIMD-8 & $4.50 \times 10^{-3}$  & SIMD-8 & $4.50 \times 10^{-3}$    \\ \hline
Load Unit               & default config & $1.56 \times 10^{-4}$  & default config & $1.56 \times 10^{-4}$    \\ \hline
Store Unit              & default config & $4.89 \times 10^{-4}$  & default config & $4.89 \times 10^{-4}$    \\ \hline
Flow Unit               & default config & $4.30 \times 10^{-4}$  & default config & $4.30 \times 10^{-4}$    \\ \hline
Instruction RAM Module  & 8 Banks; Unit-Level Concurrency & $3.68 \times 10^{-2}$  & 1 Bank; No Unit-Level Concurrency & $4.45 \times 10^{-3}$   \\ \hline
Operand RAM Module      & 16 Banks; Unit-Level Concurrency & $7.86 \times 10^{-2}$  & 3 Banks; No Unit-Level Concurrency & $1.41 \times 10^{-2}$   \\ \hline
NoC Routers             & 128-bit Wide Buffers for Data Routers & $1.36 \times 10^{-2}$  & 32-bit Wide Buffers for Data Routers & $0.71 \times 10^{-2}$   \\ \hline
Total PE Size               & \multicolumn{2}{|c|}{$0.150mm^2$} &                 \multicolumn{2}{|c|}{$0.032mm^2$}     \\ \hline
\end{tabular}
\end{table*}

\subsection{Use Scenarios for RISC-NN}\label{sec:discuss:hardware}

Currently \emph{RISC-NN} can be used either as a library-based \emph{NN} accelerator or as an energy-efficient alternative to FPGA-based \emph{NN} accelerator, but our ultimate goal is to make it an automatic end platform for high-level \emph{NN} frameworks:
\begin{compactitem}
\item\textbf{As a Library-Based NN Accelerator.} As is discussed in Section \ref{sec:scheme:programming-model}, the current version of \emph{RISC-NN} uses a semi-automatic programming model which still needs some effort from programmers to generate performance and energy efficient implementations. Though not ideal, this is an acceptable overhead given the fact that only a few \emph{NN} applications will be used by a certain customer and applications themselves are not changed frequently. As a result, \emph{RISC-NN} service providers can develop libraries of typical \emph{NN} applications for end users to use. Such library-based use scenarios are widely adopted in \emph{NN} accelerators \cite{TPU,GAP-8,PULP-NN} and proven to be successful.
\item\textbf{As an Energy Efficient Alternative to FPGA-based NN Accelerators.} In previous sections, we proves that \emph{RISC-NN} has achieved high flexibility in \emph{NN} optimization implementation, similar to FPGA-based approaches. Meanwhile, \emph{RISC-NN} can achieve significantly better energy efficiency, making it an energy efficient alternative to FPGA-based approaches.
\item\textbf{As an Automatic End Platform for High-Level NN Framework.} Our future work is to develop a compiler to automatically translate the codes of high-level \emph{NN} frameworks (e.g., PyTorch and Tensorflow) into optimized instructions for \emph{RISC-NN}. Given the static nature of \emph{NN} applications and the rich prior research of RISC compiler optimizations, we believe such a goal is achievable.
\end{compactitem}

\subsection{Hardware Design Space Exploration}\label{sec:discuss:design-space}

Though \emph{RISC-NN} is mainly designed to support cloud-based \emph{NN} acceleration which has a strict performance and energy efficiency requirement, we believe it can also fit into other markets with some design adaptions. 

Without loss of generality, we discuss how to adapt \emph{RISC-NN} to fit the requirements of \emph{IoT} (Internet-of-Things) \emph{NN} applications, which have a strict power/area budget to meet the target performance. Table \ref{table:discuss:area-compare} shows the details of such an adaption (i.e., \emph{IoT RISC-NN}). Compared with \emph{RISC-NN}, \emph{IoT RISC-NN} manages to achieve similar energy efficiency but reduce 78.6\% of the PE area by slashing the number of RAM banks, router buffer depth and concurrency. This comes at a cost of reduction in available performance--\emph{IoT RISC-NN} only achieves 7\% of MAC Unit Utilization with \texttt{AlexNet\_CONV2} compared with 81\% in \emph{RISC-NN}. Yet, \emph{IoT RISC-NN} still achieves a performance of 135.2GOPS which is well above the performance requirements of typical \emph{IoT NN} applications.


\section{Conclusions} \label{sec:conclude}

We believe RISC is a better design choice for \emph{NN} accelerators over CISC. To prove our concept, we propose a many-core RISC \emph{NN} accelerator (i.e., \emph{RISC-NN}) with a 11-instruction Very-RISC ISA and other carefully selected design solutions such as distributed control, dataflow and SIMD.  

We show that \emph{RISC-NN} can realize all the necessary instructions of state-of-the-art \emph{CISC NN Accelerators} \cite{Cambricon,TPU}.
As proof of its superior programmability, our experiments show that \emph{RISC-NN} can realize the advanced \emph{CNN} optimizations and optimization for \emph{Sparse NN}, both of which previously were only available in \emph{ASIC NN Accelerators} \cite{Eyeriss}\cite{Cambricon-X}. For a variety of \emph{NN} applications, \emph{RISC-NN} achieves on average 11.88$\times$ performance efficiency over state-of-the-art \emph{Nvidia Titan Xp GPGPU}. It also achieves on average 1.29$\times$, 8.37$\times$ and 21.71$\times$ performance efficiency over CISC-based TPU in \emph{CNN}, \emph{MLP} and \emph{LSTM} applications. Finally, by applying pruning optimization, \emph{RISC-NN} achieves additional on average 26.06\% performance improvement and 33.13\% energy reduction for \emph{Sparse NN} applications.

\bibliographystyle{abbrv}
\bibliography{ref}
\end{document}